\documentclass[a4paper]{article}

\usepackage[english]{babel}
\usepackage[utf8x]{inputenc}
\usepackage[T1]{fontenc}

\usepackage[a4paper,top=3cm,bottom=2cm,left=3cm,right=3cm,marginparwidth=2cm]{geometry}

\usepackage{lineno}

\usepackage{amsmath}
\usepackage{graphicx}
\usepackage{authblk}
\usepackage{url}
\usepackage[colorinlistoftodos]{todonotes}
\usepackage[colorlinks=true, allcolors=blue]{hyperref}
\def\bit{\begin{itemize}}
\def\eit{\end{itemize}}
\def\beqr{\begin{eqnarray}}
\def\eeqr{\end{eqnarray}}
\def\bfig{\begin{figure}}
\def\efig{\end{figure}}
\def\gm{\gamma}

\def\eps{\epsilon}

\def\bt{\beta}
\def\eps{\epsilon}

\def\sg{\sigma}

\def\fr{\frac}

\def\CCC{C$^{3}$}
\numberwithin{equation}{subsection}

\newcommand\snowmass{\begin{center}\rule[-0.2in]{\hsize}{0.01in}\\\rule{\hsize}{0.01in}\\
\vskip 0.1in Submitted to the  Proceedings of the US Community Study\\ 
on the Future of Particle Physics (Snowmass 2021)\\ 
\rule{\hsize}{0.01in}\\\rule[+0.2in]{\hsize}{0.01in} \end{center}}

\begin{document}

\title{\textbf{Future Collider Options for the US}}

\author{
\parbox{\linewidth}{\centering
P.~C.~Bhat\footnote{email:pushpa@fnal.gov}, S.~Jindariani\footnote{email:sergo@fnal.gov}, G.~Ambrosio, G.~Apollinari, S.~Belomestnykh, A.~Bross, J.~Butler, A.~Canepa, D.~Elvira, P.~Fox, Z.~Gecse, E.~Gianfelice-Wendt, P.~Merkel, S.~Nagaitsev, D.~Neuffer, H.~Piekarz, S.~Posen, T.~Sen, V.~Shiltsev, N.~Solyak, D.~Stratakis,  M.~Syphers,  G.~Velev, V.~Yakovlev, K.~Yonehara, A.~Zlobin}}

\affil{
Fermi National Accelerator Laboratory, Batavia, Illinois 60510, USA\\
}

%\SetWatermarkScale{4}
%\SetWatermarkLightness{0.3}
\maketitle
\snowmass{}
\begin{abstract}
The United States has a rich history in high energy particle accelerators and colliders -- both lepton and hadron machines, which have enabled several major discoveries in elementary particle physics. To ensure continued progress in the field, U.S. leadership as a key partner in building next generation collider facilities abroad is essential; also  critically important is the exploring of options to host a future collider in the U.S. The "Snowmass" study and the subsequent Particle Physics Project Prioritization Panel (P5) process provide the timely opportunity to develop strategies  for both. What we do now will shape the future of our field and whether the U.S.  will remain a world leader in these areas.  In this white paper, we briefly discuss the US engagement in proposed collider projects abroad and describe future collider options for the U.S. We also call for initiating an integrated R\&D program for future colliders.
\end{abstract}
\newpage
\tableofcontents
%\begin{linenumbers}
\newpage
\section{Executive Summary}
\label{sec:ES}
%[PB]

World-leading global accelerator facilities require decades of planning, preparation, research and development (R\&D), construction and commissioning. Studies and planning should be undertaken now to prepare for major particle physics projects for beyond the time horizon of current major projects   in the U.S., i.e., from the 2030s into mid-century.   As the U.S. community participates in the planning process and strategy development abroad, this Snowmass is a great opportunity for the world HEP community to consider options for major complementary facilities that can be hosted in the United States. 
 
In 2014, following the previous Snowmass study, the P5 recommended a strong program for Building for Discovery in the U.S., while supporting the LHC program as a top priority and encouraging participation in the ILC.  Implementation of that program in the U.S., the major component of which is the Long Baseline Neutrino Facility (LBNF) with the Deep Underground Neutrino Experiment (DUNE), is now underway. To produce ultra-intense neutrino beams for DUNE,  construction of a new 800 MeV Superconducting Radio Frequency (SRF) PIP-II (Proton Improvement Plan-II) accelerator is being constructed.  These projects are expected to be completed by the end of this decade.  In parallel, the U.S. collider community is engaged in physics at the LHC experiments and in the upgrades for the HL-LHC at CERN, which will commence later in this decade.

After the remarkable discovery of the Higgs boson at the LHC in 2012, a global consensus for an $e^+e^-$ Higgs factory as the next collider developed. The proposal for consideration of hosting the International Linear Collider (ILC) ~\cite{ILC_TDR-v3-I, ILC_TDR-v3-II} in Japan has now been on the table for a decade ~\cite{BhatTaylorNaturePhys}.  While the international community has been studying the SRF-based ILC for decades, and significant progress has been made in accelerator technology, a decision on hosting the ILC in Japan  still seems distant.  Recent indications are that the Japanese government deems it still premature to make such a decision.  If the ILC "Higgs Factory" has to be realized in a timely fashion, this Snowmass study and P5 should consider the option of hosting it in the U.S.  The conditions under which the ILC was considered in the U.S. previously and the estimated costs are quite different from the current situation. It is the most mature technology for a linear collider, ready for construction.  It is worthwhile to revisit this option in the U.S.

Several other attractive future collider concepts which are potentially feasible to be constructed in the U.S., in particular, those which can be built at Fermilab, are being considered during this Snowmass study.  These are intermediate-scale and compact collider projects that could prove to be cost-effective and timely, and help advance particle physics beyond the HL-LHC goals.   These options include 
\begin{itemize}
    \item a novel ``Cool Copper Collider (C$^3$)" linear collider concept (250 GeV to potentially 550~GeV collider can fit on Fermilab site)
    \item linear colliders based on high gradient SRF (in the range of 50 MV/m to 90 MV/m; standing wave or travelling wave structures; 250 -- 500 GeV facility at Fermilab).
    \item 16-km circumference site-filler circular $e^+e^-$ collider, from Z to the Higgs (90 -- 240~GeV)
    \item muon colliders from Higgs Factory (125 GeV) to a maximum energy of 8 -- 10 TeV, in three or four stages 
    \item a proton-proton collider (24 -- 27 TeV)  in a 16 km circumference site-filler tunnel.
\end{itemize}

 An $e^+e^-$ Higgs factory is of immense and immediate interest to the global community, and, therefore, we discuss promising options for the U.S.  In particular, in this white paper, we propose siting options for a linear collider Higgs Factory at Fermilab. These options, linear $e^+e^-$ colliders based on normal conducting RF operating at liquid nitrogen temperatures (as in the case of C$^3$) or those based on high gradient SRF, could provide instantaneous luminosity similar to that advertised for the 250 GeV ILC and be upgradeable to higher collision energies.  The Fermilab site filler circular $e^+e^-$ collider also could provide comparable luminosity of $\geq 10^{34}$ cm$^{-2}$s$^{-1}$.  These options should be studied rigorously and  the required R\&D should be carried out in the coming years to potentially realize one of these machines in the U.S.  
 
The U.S. is ramping up its engagement in the efforts on Future Circular Colliders (FCC) at CERN. The European community, led by CERN, is now carrying out technical and financial feasibility studies, for a $\simeq 90$ km circular tunnel  in the Geneva area that would house an $e^+e^-$ circular collider (FCC-ee), to operate at $\sqrt s = 90$~GeV to 365 GeV (Z pole to Higgs and above $t\bar{t}$ threshold), to be followed, after the completion of its mission, by a hadron collider in the same tunnel (FCC-hh). Apart from the pioneering R\&D  on the high $Q_0$ SRF and high field (HF) magnets that are needed for the FCC-ee, and FCC-hh, respectively, new US-CERN collaborative efforts on tunneling issues, civil engineering, accelerator design, beam physics, etc. are developing.  A US DOE-CERN agreement was signed in December 2020 to formalize collaborations in the FCC efforts.  

An international muon collider collaboration (IMCC) has been formed, initiated by CERN, due to a resurgence of interest in the HEP community and as an option for the future of CERN.  A multi-TeV muon collider, would be both a precision and a discovery machine, providing excellent precision for Higgs coupling measurements and great direct reach for new physics. The muon collider technology is challenging, providing unique opportunities for innovation.  The machine can be staged and operated to achieve important physics goals at each stage. The US community is engaged  with the IMCC in the ongoing muon collider studies.
\par
A phased Muon Collider project starting with a demonstrator and necessary R\&D, followed by 125 GeV, 600 GeV, and later upgrades to a multi-TeV muon collider occurring in two stages (2--3 TeV and then 8--10 TeV) is possible on the Fermilab site. The scenario is cost-efficient and has a high potential for physics at each stage. This paper discusses technical aspects for each stage and outlines a roadmap for the technology demonstration and construction.
\par
This paper also discusses a proton-proton collider site filler that could operate at a collision energy in the range of 24--27 TeV. This would require $\sim$24--27 T  accelerator magnets.  If feasible, this collider would be an attractive option if FCC-ee is being constructed at CERN in the 2040s. This could later be used as an injector to a very large hadron collider in Chicagoland, if CERN takes the route to a muon collider instead of FCC.
\par
To chart a path forward for energy frontier physics, we propose that the U.S. establish an integrated future colliders R\&D program in the DOE Office of High Energy Physics (OHEP) and charge the program
\begin{itemize}
    \item to carry-out proof-of-concept technology R\&D and to develop CDR-level reports on collider options for the U.S., by the time of the next Snowmass and P5 (ca. 2029),
    \item to develop synergistic engagement in projects proposed abroad (FCC, ILC, IMCC). 
\end{itemize}
The program should include accelerator, physics and detector studies/R\&D but particularly address major challenges for the machine options, so as to inform the decision making process by the community and the funding agencies by the end of the decade.

Major accelerator/collider facilities attract talent from all over the world, inspire young people to pursue careers in science and technology (S\&T), help contribute to tomorrow's S\&T workforce, and advance technologies which enable advanced instruments for other  sciences and industry.  The United States should undertake these projects because they not only advance HEP but also benefit the country in broad and profound ways across multiple frontiers.

\newpage
\color{black}

\newpage
\section{Introduction}
%[PB]
High energy particle accelerators and colliders have played a central role in the  experimental establishment of the Standard Model, enabling discoveries of elementary particles, extensive studies and precision measurements of their properties.  The U.S. has been at the forefront of the field, defining progress in particle physics with major discoveries, over the last sixty years. The role of U.S. leadership in advancing accelerator technology in these endeavors has also been indisputable. 

Fermilab dominated the energy frontier in particle physics research for decades, discovering three elementary fermions: the b-quark (1977), the top quark (1995) and the tau neutrino (2000). After the cancellation of the  SSC (design $\sqrt{s}$ = 40 TeV) in the U.S. in 1993, the global HEP community came together to build the Large Hadron Collider (LHC) at CERN. With the beginning of operations of the LHC in 2008, Fermilab switched its focus to flagship research at the intensity frontier, and the Tevatron collider ($p\bar{p}$ at $\sqrt{s} \simeq 2$ TeV) was shutdown in 2011.  The discovery of the Higgs boson in 2012 by the CMS and ATLAS experiments at the LHC, a crowning achievement of the Standard Model and for the collider community,  illuminates the path forward.
 
 While the U.S. domestic program pursues the development and execution of neutrino and muon physics projects, the HL-LHC at CERN would provide a compelling and comprehensive program that includes essential measurements of the Higgs properties.   An $e^+e^-$ collider (either linear or circular) can provide the next outstanding opportunity to investigate the properties of the Higgs boson, a unique and special particle in the SM, in detail and with exquisite precision.  Beyond an $e^+e^-$ collider for studies of the Higgs, either a very high energy, post-LHC proton-proton collider or a multi-TeV muon collider would provide extensive direct reach for new physics beyond the SM.
 \par
The U.S. particle physics community study, "Snowmass" and the Particle Physics Project Prioritization Panel (P5) process, provide the context and opportunity for in-depth studies of these future collider facility options.  Several white papers have been written for this Snowmass Study advocating consideration of various global collider options as well as those that are suitable for hosting in the U.S. We have made an attempt here to discuss some new proposals and briefly discuss other options described in detail elsewhere and cite the relevant papers. To advance these studies and planning, Fermilab recently launched a Future Colliders initiative, and in collaboration with the Snowmass energy and accelerator frontiers, a series of Agora events are being held to facilitate discussion of various collider options.  These discussions have also provided valuable input for this white paper.
 
    \label{sec:Intro}
\subsection{Physics Landscape\label{sec:physics}}
%[SJ, PB]
The Standard Model (SM) developed in the 1960's and 1970's describes a universe in which fermions, the fundamental constituents of matter, interact via fundamental forces propagated by gauge bosons. This description of elementary particles and their interactions has been validated extensively through precision experiments and found to be incredibly successful at describing our world. The discovery of a Higgs boson \cite{ATLAS:2012yve,CMS:2012qbp}, which was the last missing piece of the Standard Model, was another major triumph.
 
However, despite the huge success of the SM, there are a number of experimental observations that it fails to explain. It does not fully explain the baryon asymmetry, incorporate the theory of gravitation as described by general relativity, or account for the accelerating expansion of the Universe as possibly described by dark energy. The model does not contain any viable dark matter particle that possesses all of the required properties deduced from cosmology and astrophysics. It also does not incorporate neutrino oscillations and their non-zero masses. Furthermore, the model suffers from several internal shortcomings, such as the hierarchy problem, where fine-tuned cancellations of large quantum corrections are required in order for the Higgs boson mass to be near the electroweak scale. It is evident that the Standard Model is just an effective theory that appears, so far, to be valid at the energies experimentally accessible today.

For the next two decades, the LHC will remain the highest energy collider in the world. The full LHC dataset is expected to be approximately 3 ab$^{-1}$, 20 times more than what we have today.  Such a dataset will provide great opportunities for studies of the SM, including detailed characterization of the Higgs boson. Deviations from Standard Model predictions in these measurements can be an indirect evidence of new physics at energy scales higher than those accessible directly. Precision on many of the Higgs boson couplings at the HL-LHC is expected to reach few percent level, thus allowing to probe large phase space of new physics. Besides detailed exploration of the SM, the LHC is a discovery machine. LHC data will greatly extend the sensitivity for new physics, with excellent chances for fundamental discoveries. 

However, it is conceivable that the HL-LHC dataset will not be sufficient to discover and fully characterize new physics. Higher collision energies would enable exploration of the laws of Nature at ever-shorter distances, providing a deeper understanding of fundamental particles and fields. Furthermore, both hadron and lepton future colliders \cite{shiltsev2021modern} enable even more precise measurements of Standard Model parameters, including those in the Higgs sector, which in turn provide deeper insight into the mechanism of electroweak symmetry breaking. 

It is evident that detailed exploration of the electroweak sector of the Standard Model remains a high priority. This includes precise determination of the nature of the Higgs boson, including measurements of its properties and couplings. In particular, measuring Higgs boson couplings at the sub-percent level allows to constrain a wide range of new physics models or provide first indirect evidence of beyond the SM (BSM) particles or forces. Measurements of the Higgs boson decay rate to invisible particles and its total width are also very important for discovering or constraining BSM physics. Beyond the couplings, measurements of Higgs boson self-interactions (both trilinear and quartic) allow to fully establish the shape of the Higgs potential and verify if it agrees with the SM predictions.  While the Higgs boson remains the centerpiece for the precision program, many other rare SM processes continue to attract significant interest. These include studies of lepton flavor universality in B meson decays, flavor changing neutral currents in top decays, $\tau\rightarrow 3\mu$ and others. Measurements of the mass and width of the vector bosons, the electroweak mixing angle, and the vector boson scattering amplitudes would further shed light on the underlying structure of the electroweak sector of the SM.

Increasing the energy scales accessible at the colliders allows the laws of nature to be probed directly at ever shorter distances, which permits the exploration of underlying principles that may govern the properties of the elementary fields. It may lead to the discovery of new particles or forces that are impossible to produce, or simply not sufficiently abundant, at present colliders. The purest science driver is therefore the exploration of the unknown. Prominent targets include particle explanations of dark matter or the matter-antimatter asymmetry, probes for the existence of new gauge or space-time symmetries, as well as tests of theories containing multi-TeV resonances. Furthermore, only higher-energy colliders may probe the key question of whether the particles currently considered elementary are composite states at shorter distances. Finally, the future colliders program has certain unique synergies with the neutrino and precision frontiers, which enable a complementary program of physics measurements at neutrino factories and/or fixed-target experiments.
\subsection{Existing and proposed facilities}
%[PB]
 Following the recommendation of the U.S. P5 in 2014, a strong program for "Building for Discovery" in the U.S. for neutrino and muon-beam based physics is underway.   The major component of the program, the Long Baseline Neutrino Facility (LBNF) to host the Deep Underground Neutrino Experiment (DUNE), is being implemented.  A new 800 MeV Superconducting Radio Frequency (SRF) PIP-II (Proton Improvement Plan-II) accelerator under construction, will provide ultra-intense neutrino beams to DUNE.  The LBNF/DUNE and PIP-II accelerator projects are expected to be completed by the end of this decade.  
 
 The U.S. collider physics community is engaged in physics and upgrades at the LHC, including the HL-LHC program at CERN which will commence later in this decade.  The U.S. is engaged in both accelerator upgrades and upgrades of the LHC experiments to ensure maximum physics output from the LHC program. 
  
 The Higgs boson discovery in 2012 at the LHC ~\cite{ATLAS:2012yve,CMS:2012qbp} completed the Standard Model (SM), but also led to a greatly renewed interest in the world HEP community towards planning next generation colliders \cite{shiltsev2021modern}.   The need for two categories of colliders is apparent: 1) a Higgs Factory that would enable extensive and precision studies of the Higgs boson; and 2) a post-LHC, $\sim100$~TeV scale, hadron collider to advance the energy frontier explorations in search of new physics beyond the Standard Model (BSM).   

Since the measured Higgs mass is $\sim125$ GeV, several proposals for an electron–positron Higgs Factory, have been made in Europe and Asia: 
\begin{itemize}
    \item International Linear Collider (ILC) (being considered in Japan), 
    \item Compact Linear Collider (CLIC) at CERN,
    \item Future Circular Collider (FCC-ee) to be followed by FCC-hh at CERN, and
    \item Circular Electron–Positron Collider (CEPC) to be followed by SppC in China. 
\end{itemize}

 Recently, there has been a significant resurgence of interest in muon colliders, which have also been studied for over two decades.  A muon collider could be built as a Higgs Factory at $\sqrt{s}$ of $\sim 125$~GeV for precision studies of the Higgs properties while multi-TeV muon colliders could provide competitive discovery potential and precision measurements, on par with hadron colliders at several tens of TeV.  
 
 Apart from the aforementioned global collider projects under development over the past couple of decades, there are many novel concepts for colliders of modest size and cost, that have emerged in the past couple of years.

\subsection{Emerging Concepts and proposals}
%[PB]
The energy frontier facilities that address the HEP mission of studying the Higgs boson in detail and with great precision, and for pursuing new physics beyond the HL-LHC reach, include linear $e^+e^-$ colliders,  circular (preferably large circumference) $e^+e^-$ storage rings, muon colliders, and high energy hadron colliders. We have mentioned global megaprojects of ILC, FCC, CLIC, CEPC/SppC that have been studied extensively and we will  discuss the U.S. engagement in some of these projects very briefly in this paper. With resurgence of interest in a muon collider, an international muon collaboration (IMCC) has been formed. The U.S. engagement in IMCC will also be discussed.  In addition to the robust machine proposals mentioned in the previous section, ideas for intermediate scale, modest-cost, compact colliders have emerged recently.  These proposal include:  

\begin{itemize}
    \item a novel ``Cool Copper Collider (C$^3$)" linear collider concept (250~GeV to potentially 550~GeV collider can fit on Fermilab site)
    \item linear colliders based on high gradient SRF (in the accelerating gradient range of 50~MV/m to 90~MV/m; standing wave or travelling wave structures). A center of mass energy reach between  250 and 500 GeV with the facility's central campus within the Fermilab site is possible.
    \item 16-km circumference site-filler circular $e^+e^-$ collider, from Z to the Higgs (90 -- 240~GeV)
    \item muon colliders from Higgs Factory (125 GeV) to a maximum energy of 8 -- 10~TeV, in three or four stages 
    \item a proton-proton collider (24 -- 27 TeV)  in a 16 km circumference site-filler tunnel
\end{itemize}

Some of these have been described in detail elsewhere; we briefly outline them here and cite relevant papers.  Other proposals we discuss in some detail.

We would like to emphasize that a strong R\&D program addressing major challenges for these concepts need to be undertaken to make timely progress.  Early emphasis in the R\&D could be placed on design/simulation studies including tools development that would have applicability for all of the promising collider concepts.  Focused and intense R\&D on most promising collider option(s) should be undertaken over the next several years to investigate and address major technological challenges, perform preliminary feasibility studies and produce CDR-level reports before the start of the next Snowmass study.  Therefore, we call for funding for an integrated collider R\&D program to be supported by the DOE office of High Energy Physics. Synergies with intensity frontier facility requirements, where available, should be taken into consideration. 
\section{US Engagement in Global HEP Projects}
\label{sec:globalprojects}

\subsection{ILC}
%[Sergey B, Sam, Sergei N, PB]

\subsubsection{Introduction and status}

 The International Linear Collider (ILC) has been the prime candidate for a Higgs Factory since the discovery of the Higgs boson. It is under consideration to be hosted in Japan.  The collider facility will be about 20.5 kilometers in total length. ILC will accelerate beams of electrons and positrons to 125 GeV each in two superconducting RF linacs, and collide them at the center of the machine where detector(s) will record the data from the collisions, see e.g., \cite{aihara2019international, bambade2019international, ILCSnowmass2021}.

Operating at 250 GeV, the ILC (referred to as ILC250) will provide for copious production of the Higgs boson along with a Z boson via the process $e^{+}e^{-}\rightarrow ZH$. The design instantaneous luminosity of the ILC250 will be $1.35\times10^{34}$ cm$^{-2}$s$^{-1}$. With beam polarization (80\% for electrons and 30\% for positrons), the effective luminosity would be about $2.0\times10^{35}$ cm$^{-2}$s$^{-1}$) \cite{padamsee2019impact}. There are  proposals to upgrade to higher luminosity (up to $8.1\times10^{34}$ cm$^{-2}$s$^{-1}$).  With some modest investment, the ILC will be upgradeable to higher collision energies up to 380 GeV in the future.  In principle, upgrades to 500 GeV, 1 TeV, and beyond are possible \cite{padamsee2021ilc, ILCSnowmass2021}.

The underlying SRF linac technology (originally developed for TESLA collider project \cite{TESLAcavity}) is mature and has been utilized in a number of SRF projects throughout the world, such as free electron laser facilities European XFEL at DESY and LCLS-II at SLAC. The cavity production data (from 831 cavities) for European XFEL  show that it is possible to mass-produce cavities with desired gradient and efficiency. 

As reported in the ILC TDR \cite{ILC_TDR-v3-I}, during phase II of the R\&D program, the $(94 \pm 6)$\% yield has been achieved for cavities that demonstrated accelerating gradients >28~MV/m and $(75\pm11$\% for 35~MV/m (ILC specification 31.5 MV/m). This ensemble of cavities has an average gradient of 37.1~MV/m. The yields were demonstrated after re-treating cavities with gradients outside the ILC specification. Laboratories from three regions -- America, Asia, and Europe -- developed this critical technology over the years. Cryomodules are built globally at DESY, CEA, FNAL, JLAB, KEK, and in China. Cryomodules meeting the ILC gradient specifications have demonstrated operation with beam at Fermilab \cite{FASTrecord} and KEK \cite{SRF_2021}.

SRF was chosen as the ILC technology in 2005 for multiple reasons, including:
\begin{itemize}
    \item power-efficient acceleration (high beam power to AC power efficiency) with the total AC power of $\sim$110~MW for ILC250
    \item relaxed tolerances compared to room-temperature designs due to larger apertures
    \item larger vertical beam spot at collision (7.7~nm) than for normal conducting linear colliders
    \item due to low RF losses, RF pulse length and bunch separation (727 $\mu$s and 554~ns) are large enough to allow corrections between pulses as well as within a bunch train (intra-train feedback)
    \item luminosity upgrades via increased beam power
    \item energy upgrades with gradient advances in SRF technology
\end{itemize}

Other critical items for ILC accelerator technologies are nano-beams for final focus, low-emittance damping rings, and positron production. Accelerator Test Facility 2 (ATF2) was built at KEK in 2008 as a test-bench for the ILC final focus scheme. The primary goals were to achieve a 37 nm vertical beam size at the interaction point (IP), and to demonstrate beam stabilization at the nm level. After scaling for the beam energies from 1.3 GeV (ATF2) to 250 GeV, the 37 nm beam size corresponds to the TDR design value of 5.7 nm at 250 GeV beam energy. The goal has been reached within 10\% validating the final focus design. Experiments at CESR-TA (CESR Test Accelerator) at Cornell have demonstrated confidence in the ILC damping ring parameters.

The baseline machine parameters remain stable since the publication of the Technical Design Report (TDR) in 2013 \cite{ILC_TDR-v3-I, ILC_TDR-v3-II} with some recent updates \cite{evans2017international, bambade2019international, ILCSnowmass2021}. The ILC cost was evaluated in 2012 for TDR using a detailed, bottoms-up approach. The cost covers accelerator construction over 9 years plus 1 year commissioning. It includes fabrication, procurement, testing, installation, and commissioning of the whole accelerator, tunnels, buildings etc., and operation of central laboratory. It does not cover costs during the preparation phase, design work, land acquisition, infrastructure. The overall cost of ILC250 is in the range 4.8 -- 5.3 BILCU  (Billions of ILC Units, 1~ILCU is approximately 1~US\$), and does not include labor and detectors. The labor is evaluated at 10,000 person-years. The detectors cost is 0.7 BILC plus 2,200 person-years. 

\subsubsection{US/Fermilab Engagement}

The U.S. institutions have been involved in the development of the SRF TESLA technology from the very beginning, making important contributions. In fact, the first TESLA collider workshop was held in the USA, at Cornell University \cite{1st_TESLA_ws} in 1990. In addition to SRF, the U.S. laboratories have participated in almost all other aspects of the ILC development: electron and positron sources, RF power distribution, damping rings, beam delivery system, beam dynamics, instrumentation, detector R\&D. Fermilab, in particular, contributed to developing fundamental RF power couplers, cavity frequency tuners, the 1.3~GHz cryomodule design, design of the 3.9~GHz cryomodule and all its components, etc. 

In recent years, the U.S. community has been engaged in collaboration with Japan in the framework of the ILC Cost Reduction R\&D Program and more generally in updating ILC plans via participation in the ILC International Development Team (IDT) \cite{ICFA_IDT}. New surface treatment processes were developed at Fermilab for the cavity preparation process,  allowing the cavities to achieve higher accelerating gradients while improving the quality factors at the same time. Applying these new treatments to ILC would provide opportunities to i) improve the efficiency of ILC250, ii) upgrade the luminosity, and iii) upgrade the energy of collisions as described in \cite{padamsee2019impact}. The anticipated saving from the ILC Cost Reduction R\&D is $\sim10$~\%. In addition to the cavity R\&D, Fermilab scientists and engineers are involved in updating designs of the ILC cryomodule and some components, efforts to harmonize pressure vessel codes across the three regions, and developing new SRF crab cavities.

%\subsubsection{Future US engagement}
As of this writing, the plan for the U.S. community is to continue engagement in preparations for the ILC in Japan. If the efforts led by Japan continues, it is anticipated that after a couple of years of transition period with very modest investment in the most critical, high priority activities, an approximately four-year Pre-Lab will be organized that would prepare the project for the beginning of construction.

\subsection{FCC}
%[Sergey B, Sergei N, Vladimir]

\subsubsection{Introduction and status}
The proposed circular collider FCC-ee is a well-studied $e^+e^-$ machine to be located surrounding CERN and Geneva. The double-ring collider would operate at center of mass energies ranging from the $Z$-pole (91 GeV) to $t \bar{t}$ (365 GeV). The present optimized main tunnel length is 91.2 km. Bunched beams (with $\sim$~ampere current) maintained by SRF cavities would circulate in the two rings, one per beam, and collide in up to four interaction regions. The projected luminosity per IP ranges from $1.8\times 10^{36}$ cm$^{-2}$s$^{-1}$ at the $Z$ to $1.25\times 10^{34}$ cm$^{-2}$s$^{-1}$ at the $t \bar{t}$ within the limit of 50~MW of synchrotron radiation power loss per beam. A full-energy injector ring located in the same tunnel would top-up the beam currents in the collider rings. In addition to the new ring, the injector chain would reuse significant parts of the present CERN infrastructure. A CDR has been written in 2018 \cite{fcc2019fcc} and recently updated to a 4-IP lattice. Significant design efforts and R\&D have been completed including lattice, magnets, IR, site, and staging. The crucial future technical R\&D will concentrate on developing the 7.7 GV SRF systems, which would include higher order mode (HOM) damped cavities and highly efficient RF klystrons. 

Though technically the project is nearly ready to proceed, it needs to wait for the HL-LHC operational program to be completed leading to a start date for the FCC-ee of around 2042. Its construction cost is projected by the proponents to be about 10.5 BCHF (European accounting) and additional 1.1 BCHF for the RF needed to go to the $t \bar{t}$ energy. The FCC collaboration carries out extensive R\&D and prototyping effort. To the project's advantage, circular $e^+e^-$ colliders overall have a half-century long history of success including CESR and PEP-II in the U.S. and LEP at CERN. Multi-ampere beams have been demonstrated at PEP-II and KEKB in Japan. The SuperKEKB collider in Tsukuba, now in operation, will demonstrate in the next few years nearly all the required accelerator physics techniques for the FCC-ee, as will the future electron ring for the electron-ion collider (EIC) at the Brookhaven National Laboratory.

Among the  main challenges for FCC-ee are: i) the peak luminosity within the given synchrotron radiation power limit $P_{SR}$ drops at higher beam energies approximately as $L \propto P_{SR}/E^3$; ii) a crab waist collision scheme with a large crossing angle, high bunch charges and mm-level vertical beta functions need solid verification; iii) SRF cavities with strong HOM damping required to support multi-ampere beams need to be tested; iv) overall cost and total facility site power reduction strategies need to be fully explored. 

Following the execution of the FCC--ee physics program, in a way similar to the hands-off between LEP/LEP2 and LHC, the FCC-ee tunnel can be dedicated to a hadronic collider called FCC-hh \cite{FCC-hh}. FCC-hh can provide proton–proton collisions with a center-of-mass energy of 100~TeV, instantaneous luminosity ranging from $5\times 10^{34}$ cm$^{-2}$s$^{-1}$ to $30\times 10^{34}$ cm$^{-2}$s$^{-1}$ and an integrated luminosity of $\sim$ 20 $ab^{−1}$ in each of the two main experiments for 25 years of operation.

The collider would use the existing CERN accelerator complex as injector facility at $\sim 3.3$ TeV from the LHC and, with a filling factor of 0.8, would require dipole fields just below 16 Tesla to keep the nominal beams on the circular orbit.
 
Many technical systems and operational concepts for FCC-hh can be scaled up from HL-LHC but will require, in some cases, additional R\&D. Particular technological challenges arise from the higher total energy in the beam (20 times that of LHC), the much increased collision debris in the experiments (40 times that of HL-LHC) and far higher levels of synchrotron radiation in the arcs (200 times that of LHC). 
 
\subsubsection{US/Fermilab Engagement}

%[Sergei N on current FCC effort]
The U.S. HEP community has long-term, very productive and close ties with CERN collider program. In general, our community supports main EPPSU'2019 recommendations to consider exploration of Higgs physics and Higgs factory as the highest priority for particle physics after completion of the LHC program. Therefore, we have to consider various options for our contributions to the FCC-ee project. There is significant expertise available in the U.S. in the area of accelerator design and corresponding R\&D, and it seems rational to establish a sub-program in the DOE OHEP which would deal with organization of the FCC-ee related effort.  

Examples of topics common to different machines include study of the machine-detector interface, beam collimation, and tuning of linear and non-linear optics. Supported topics would be a mix of theory, simulation, and hardware development and experiments.  
\subsection{Muon Collider}
\label{sec:MC}
\subsubsection{Introduction}
A colliding-beam facility based on muons has a number of advantages~\cite{Chao:2014pea}. First, since the muon is a lepton, all of the beam energy is available in the collision. Second, since the muon is roughly 200 times heavier than the electron and thus emits around $10^9$ times less synchrotron radiation than an electron beam of the same energy, it is possible to produce multi-TeV collisions in an Fermilab-sized circular collider. The large muon mass also enhances the direct “s-channel” Higgs-production rate by a factor of around 40,000 compared to that in electron–positron colliders, making it possible to scan the center-of-mass energy to measure the Higgs-boson line shape directly and to search for closely spaced states. Finally, high-energy muon colliders are the most efficient machines in terms of power per luminosity. While the above arguments are highly appealing, there are several challenges with muons. First, muons are obtained from decay of pions made by higher energy protons impinging on a target. The proton source must have a high intensity, and very efficient capture of pions is required. Second, muons have very large emittance and must be cooled quickly before their decay. Given their short time, ionization cooling~\cite{Neuffer:1983jr} is the only viable option. Moreover, conventional synchrotron accelerators are too slow and recirculating accelerators and/or pulsed synchrotrons must be considered. Because they decay while stored in the collider, muons irradiate the ring and detector with decay electrons. Shielding is essential and backgrounds will be high. 

\subsubsection{Muon Collider History}
\begin{figure}[t]
\begin{center}
\includegraphics[width=0.98\textwidth]{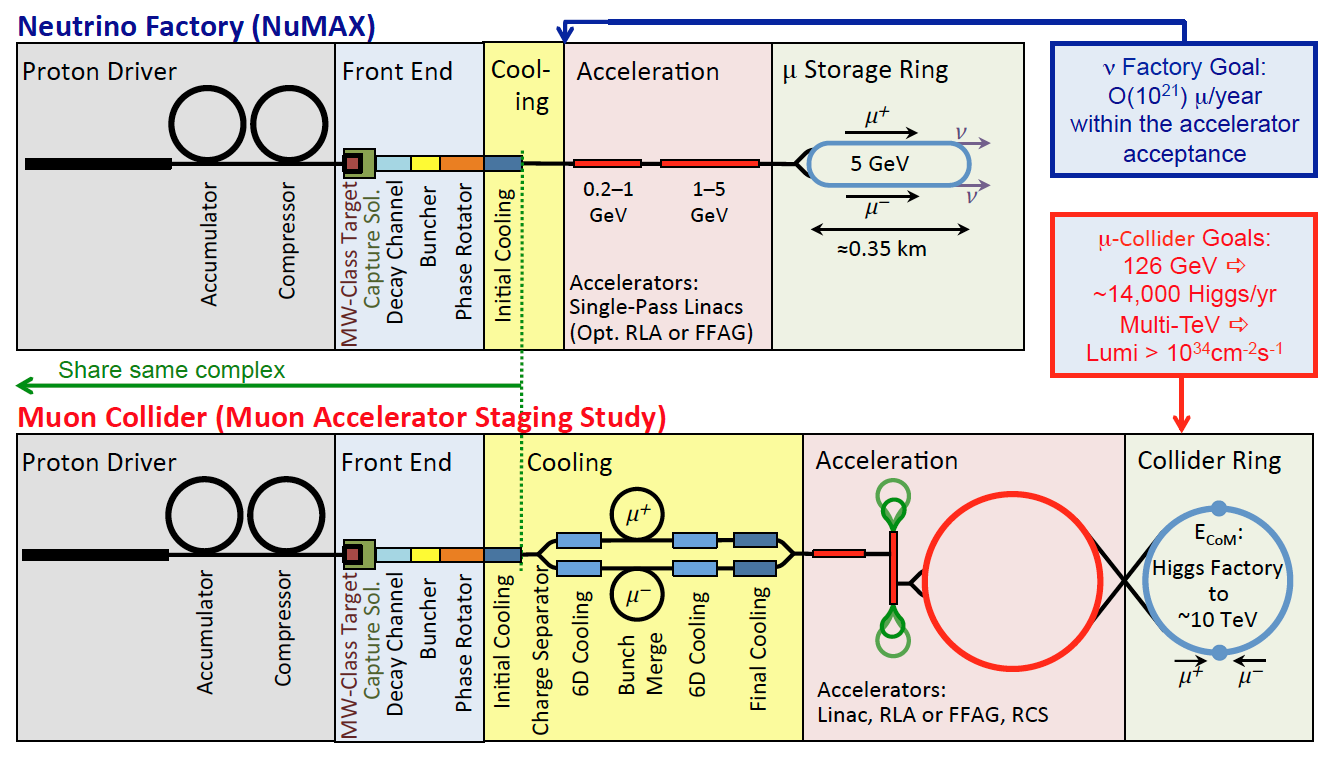}
\caption{Block diagrams showing the principal elements of a Neutrino Factory (NF) and a Muon Collider (MC).}%
\label{fig:NFMCC}
\end{center}
\end{figure}

The concept of a muon collider is not new.  Muon storage rings were mentioned in the literature in 1965~\cite{Tinlot:1965ab} and concepts for a muon collider and for the required muon cooling were developed in the 1970s and 1980s.  A muon collider collaboration was formed in the U.S. in the 1990s which delivered a design report in 1999~\cite{Ankenbrandt:1999cta}.  In 2000 the Neutrino Factory and Muon Collider Collaboration (NFMCC) was formed~\cite{Zisman:2000dn} which set out to perform a multi-year R\&D program aimed at validating the critical design concepts for the Neutrino Factory (NF) and the Muon Collider (MC). The Muon Accelerator Program (MAP)~\cite{Palmer:2013/07/02bta} was a follow-on (approved in 2011) program to the NFMCC and was tasked to assess the feasibility of the technologies required for the construction  of the NF and the MC. At the conclusion of MAP the program had produced a number of significant milestones:
\begin{enumerate}
    \item Full development of the principal elements of the NF and the MC~\cite{Palmer:2013/07/02bta} (see Figure~\ref{fig:NFMCC}).
    \item End-to-End simulation of cooling for the MC~\cite{Palmer:2016gws}.
    \item Demonstration of a mercury-jet target capable of 8~MW operation~\cite{McDonald:2010zzc}.
    \item Operation of a high-gradient 805~MHz RF cavity in high magnetic field~\cite{Bowring:2018smm}.
    \item First demonstration of muon ionization cooling (MICE~\cite{MICE:2019jkl}).
\end{enumerate}

Although MAP was terminated in 2016, work continued on documenting the program's results and has provided a ``jumping-off" point for the recently formed International Muon Collider Collaboration (section~\ref{sec:IMCC}).

\subsubsection{International Muon Collider Collaboration}
\label{sec:IMCC}
The 2019 update of the European Strategy for Particle Physics identified muon colliders as a highly promising path to reaching very high center-of-mass energies in leptonic collisions. These machines therefore combine excellent new physics discovery potential with high precision capabilities. In response to these findings, the European Laboratory Directors Group (LDG) formed a muon beam panel and charged it with delivering input the the European Accelerator R\&D Roadmap covering the development and evaluation of a muon collider option. In parallel, CERN initiated formation of a new International Muon Collider Collaboration (IMCC) to assess feasibility of building a high energy muon collider, identify critical challenges, and develop an R\&D program aimed to address them. The effort includes development of the machine-detector interface (MDI), detector concepts, and an evaluation of the physics potential. 

\begin{figure}[h]
\begin{center}
\includegraphics[width=0.98\textwidth]{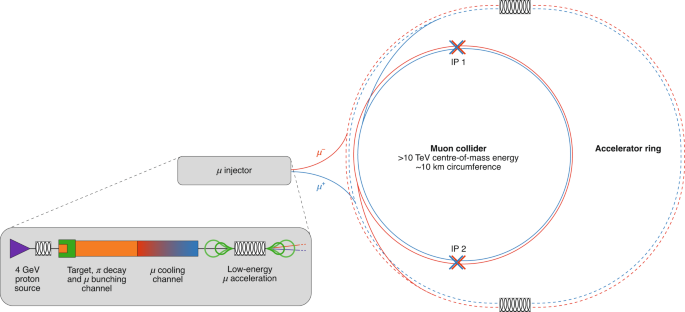}
\caption{Schematic layout of 10 TeV-class muon collider complex being study within the International Muon Collider Collaboration.  From https://muoncollider.web.cern.ch/}%
\label{fig:IMCC}
\end{center}
\end{figure}

The collaboration is hosted by CERN. The near-term goal is to establish whether an investment into a full Conceptual Design Report and a demonstrator are scientifically justified for the next European Strategy for Particle Physics Update. Depending on the outcome of this study and the decisions made at the next ESPPU, the design can be further optimised and a demonstration program can be executed in the following years. The latter contains one or more test facilities as well as the development and testing of individual components and potentially dedicated beam tests. The resulting conceptual design will demonstrate the possibility to technically commit to the collider. In this case a technical design phase will follow to prepare the approval and ultimate implementation of the collider.

The design strategy taken by IMCC relies heavily on the concepts developed by the MAP collaboration. In the baseline design, muons are produced in decays on pions produced by colliding a multi-megawatt proton beam onto a target. The muons are then cooled to the emittances necessary to achieve target luminosities, rapidly accelerated to the desired energies in order to minimize the number of muon decays, and injected into a collider ring with two interaction points. IMCC envisions a staged approach with the first stage collider operating at the center-of-mass energy of 3 TeV and the second stage at 10+ TeV (Figure~\ref{fig:IMCC}). Integrated luminosity targets per interaction point are 1 ab$^{-1}$ and 10 ab$^{-1}$, respectively. Staging allows for demonstration of performance at the lower energy and also facilitates stretching out the construction time, while executing a vibrant physics program. The front end and most of the cooling chain in the accelerator complex are common to the two stages. An alternative approach (LEMMA), which uses positrons to produce muon pairs at threshold, was also considered but had difficulties with achieving a high muon beam current and hence the necessary luminosity.

The IMCC held four "community meetings" in 2020 and 2021 to develop the scope and the plan of work to be done between now and the next ESPPU. R\&D objectives have been identified in several key areas, including muon production and cooling, neutrino induced radiation mitigation, MDI studies and optimization, and the high energy complex. Technologically, the design imposes challenging requirements on the high power targets where short proton bunch length and frequency may compromise the target's lifetime and integrity, on the high-field solenoidal magnets used in the production, collection and cooling of the muons, as well as on the specs of fast-ramping and fixed-field magnets used in the accelerator and collider rings. The ionization cooling system is a novel concept and requires careful studies for optimal integration of the absorber and RF stations inside of high magnetic fields. Successful demonstration of a partial muon cooling system is therefore crucial for the design verification. This test facility can be located at any laboratory that can provide proton beam of needed power. Currently rough dimensions of the facility have been identified and siting at CERN is being explored. Section~\ref{sec:MCdemo} describes how such a facility can be hosted at Fermilab.

\subsubsection{US Contributions to IMCC}
Despite strong interest and expertise, U.S. participation in IMCC has been mainly limited to the work done in the context of Snowmass. As mentioned above, the design strategy taken by IMCC relies heavily on the concepts developed by the MAP collaboration. The European muon beam panel included two representatives (including the co-chair) from the U.S., and a large number of scientists helped to organize the IMCC working group activities. U.S. scientists made key contributions to most areas of the IMCC design development and planning, including magnets, RF cavities, muon production and cooling, muon acceleration, beam dynamics, machine-detector interface, and the high-energy complex. Besides the accelerator design, the Energy and Theory Frontier communities in the U.S. provided strong contributions in the areas of physics studies and detector design. 

\subsubsection{Snowmass Muon Collider Forum}
In light of renewed interest in muon colliders within the United States particle physics community, the Snowmass Energy, Theory and Accelerator Frontiers have created a Muon Collider Forum. The Forum~\cite{mucollforum} meets on a monthly basis and has invited several experts to give their perspective and to educate broader community about the physics potential and technical feasibility of such a collider. In addition, it facilitates a strong bond between the particle physics community and accelerator experts and organizes related workshops. A Muon Collider Summary Report will be prepared for Snowmass to highlight key areas where U.S. can provide critical contributions to the global efforts as well as to present Fermilab as one of the options for hosting a MC in the future.    
Future U.S. contributions to the global Muon Collider R\&D roadmap are contingent on the outcome of the Snowmass and P5 processes. However, discussions within the Snowmass Muon Collider Forum started in order to identify key areas of interest and expertise, assuming that P5 will support a revival of the Muon Collider R\&D program. The areas that have been identified include design of the proton driver (in synergy with the PIP-II accelerator), targetry (in synergy with future Fermilab neutrino and precision muon programs), muon cooling design and optimization, accelerator lattice design, high-field magnet development (in synergy with the Magnet Development Program), beam acceleration using superconducting RF technology, and mitigation of the neutrino induced radiation. 
    \label{sec:FF}
\section{Linear {$e^+ e^-$} colliders at Fermilab}
\subsection{C$^3$ proposal}
%[Vladimir, SB, PB]

Cool Copper Collider (C$^3$), proposed in Ref.~\cite{nanni2021c} is based on a cold normal conducting C-band RF (NCRF) technology, which promises dramatic improvement in efficiency and breakdown rate compared to previously achieved. High accelerating gradient of 70--120~MV/m allows to reach $HZ$ production energies with a relatively small facility that could, for example, be located at the Fermilab site.  An $\sim8$-km long 250~GeV Higgs Factory (with a relatively inexpensive upgrade to 550~GeV within the same footprint) has a luminosity of $1.3\cdot10^{34}$ cm$^{-2}$s$^{-1}$ ($2.4\cdot10^{34}$ cm$^{-2}$s$^{-1}$ at 550~GeV) \cite{nanni2021c}. The estimated site power is $\sim$150~MW at 250~GeV and $\sim$175~MW at 550~GeV. In principle, C$^3$ is potentially extendable to 3 TeV by simple extension of the linac while keeping the accelerating gradient at 120 MV/m.

The key technology of C$^3$ is a structure distributing power to each accelerating cell in parallel from a common RF manifold. This allows optimization for cell efficiency (shunt impedance) while controlling peak surface electric and magnetic fields. Operation at $\sim 80$~K with liquid nitrogen cooling improves the material strength, reduces the breakdown rate, and allows higher accelerating gradients. First proof-of-principle experiments demonstrated operation up to 150~MV/m with expected robust operation up to 120~MV/m. Further R\&D in a few key areas is required (e.g., scaling modular units; developing cryogenic, cryomodule and alignment systems; integration of wakefield detuning/damping scheme into the structure design). One of the main challenges for C$^3$ is alignment and jitter. The main linac will require 5-micron structure alignment, which would be achieved by a combination of mechanical pre-alignment and beam-based alignment. A demonstration facility is proposed to support critical R\&D topics \cite{C3demo}.

While RF sources and modulators capable of powering the 250 GeV C$^3$ are commercially available, the RF source is the key cost driver for the overall cost of the machine. R\&D on reducing the RF source cost is of critical importance. The plan is to leverage significant recent developments in performance of high-power RF sources (e.g., by HEIKA collaboration). It will require significant industrialization efforts after the technology demonstration.

\begin{figure}
    \centering
    \includegraphics[width=0.8\textwidth]{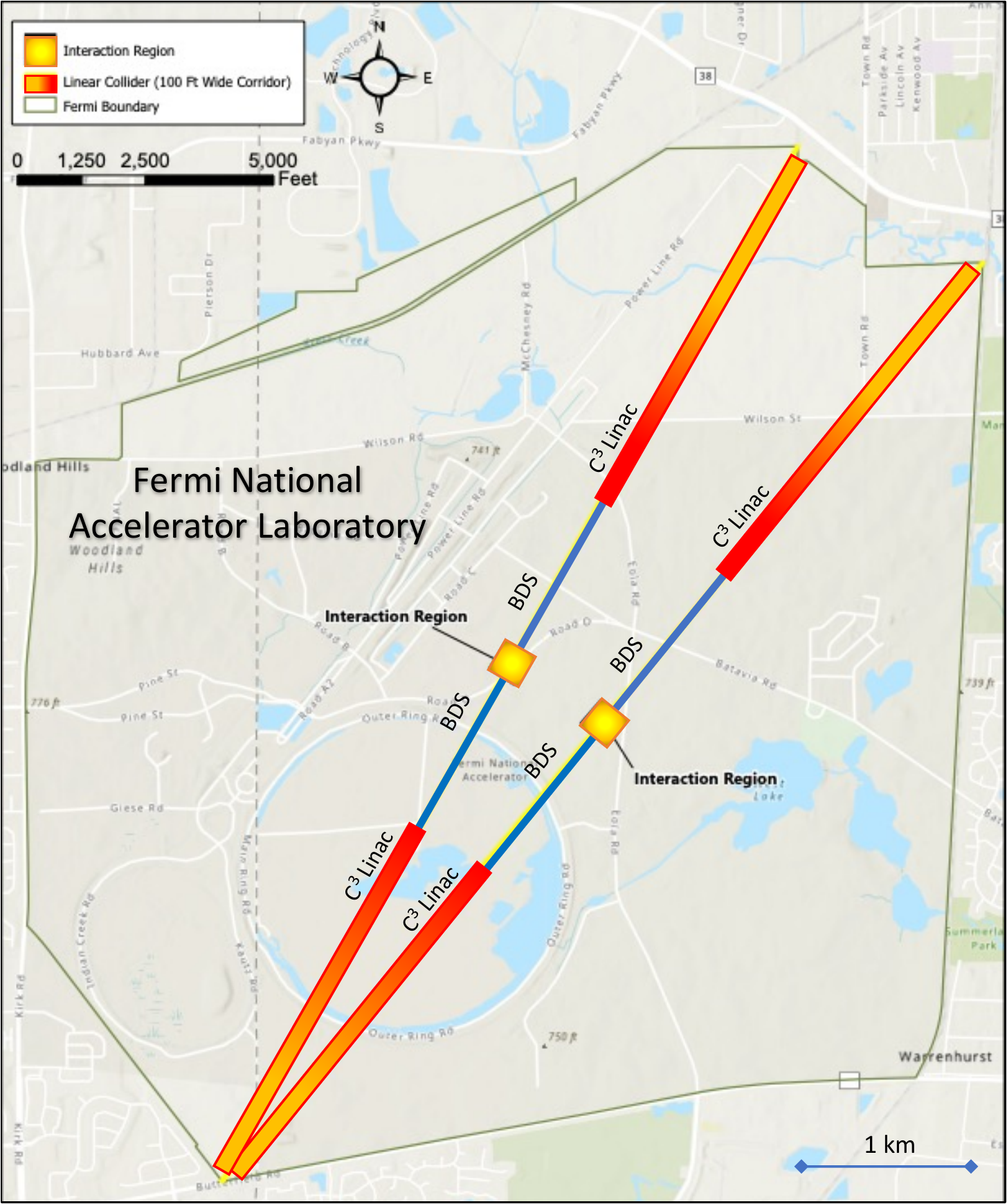}
    \caption{Possible locations for a 7-km footprint linear collider on Fermilab site considered for C$^3$.}
    \label{fig:LC_7kmLayout}
\end{figure}

\begin{figure}
    \centering
    \includegraphics[width=0.8\textwidth]{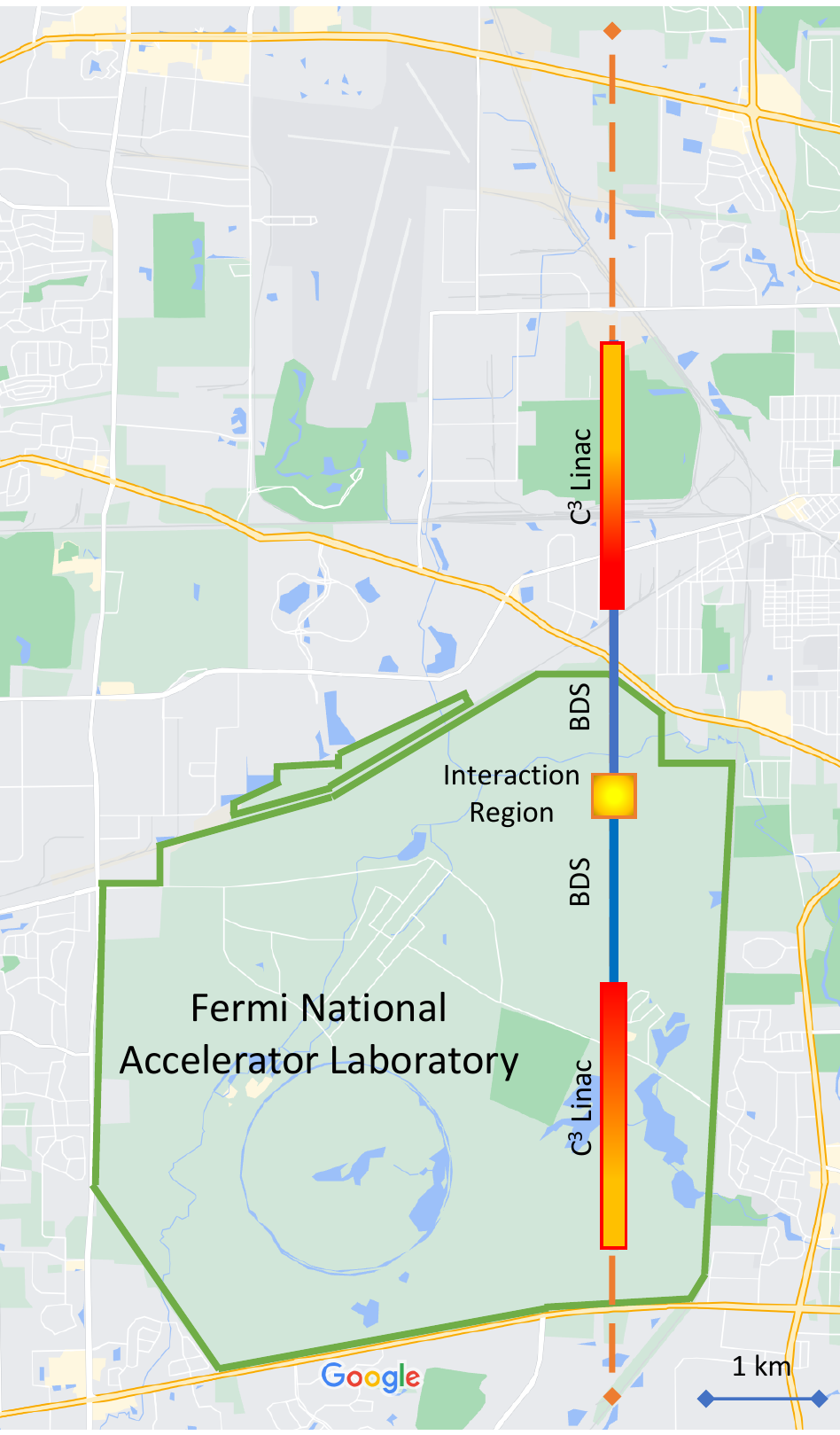}
    \caption{The 8-km footprint consisting of 5 km inside the Lab site and extending the facility under the Common Wealth Edison power company's easement, considered for C$^3$ and HELEN.}
    \label{fig:LC_8kmLayout}
\end{figure}

Fermilab site can fit a  7-km footprint linear machine entirely within its boundaries, in North East -- South West (NE--SW) orientation (See Figure~\ref{fig:LC_7kmLayout}). The 8-km footprint currently proposed to upgrade to 550 GeV, can be accommodated with about 5 km of the footprint inside the Lab site and extending the facility under the ComEd power company's easement to the north of the Lab site (North -- South (N--S) orientation). This option is shown in Figure~\ref{fig:LC_8kmLayout}. It is possible to have the machine footprint up to 12 km in this orientation and siting option, keeping the interaction region of the collider within the Lab campus.  This siting location, was, in fact, one of the options studied for the ILC at Fermilab.   Using the full 12 km length can provide upgrade paths to 750~GeV collision energy or higher.
Perhaps, further optimization of the final focus could let the 8 km machine for energy upgrade up to 550 GeV fit within the boundary of the laboratory itself, i.e., with a footprint of 7 km or less, using NE--SW orientation.     

The 8-km long C$^3$ footprint allows achieving 250~GeV center-of-mass energy with an accelerating gradient of 70~MV/m (assumed linac filling factor is 90\%). This gradient is cost-optimal for the current large-volume RF source unit cost of $\sim \$$7.5/peak-kW.  Raising the gradient to 120~MV/m would increase the energy to 550~GeV within the same footprint (a full suite of cryomodules needed for the 550 GeV operation would be installed during the 250 GeV construction, but not all of them would be powered up.) This upgrade will require development of new RF sources and/or RF pulse compression scheme. Large portions of accelerator complex are similar to other linear colliders: beam delivery system (BDS) and interaction region (IR) can be modified from the ILC design (currently C$^3$ assumes $\pm1.5$~km BDS for the 550~GeV center-of-mass energy); damping rings and injectors to be optimized with CLIC as a baseline. Costing studies so far used other linear collider estimates as inputs. The total capital cost is estimated at 3.7~BILC. The technically-driven timeline includes 2 years for a pre-demo stage, 5 years for the technology demonstration, 3 years for a string test, and 8-10 years of construction and commissioning time.

\newpage
\subsection{HELEN -- A linear collider based on advanced SRF}
%[Sergey B]

\begin{figure}
    \centering
    \includegraphics[width=0.8\textwidth]{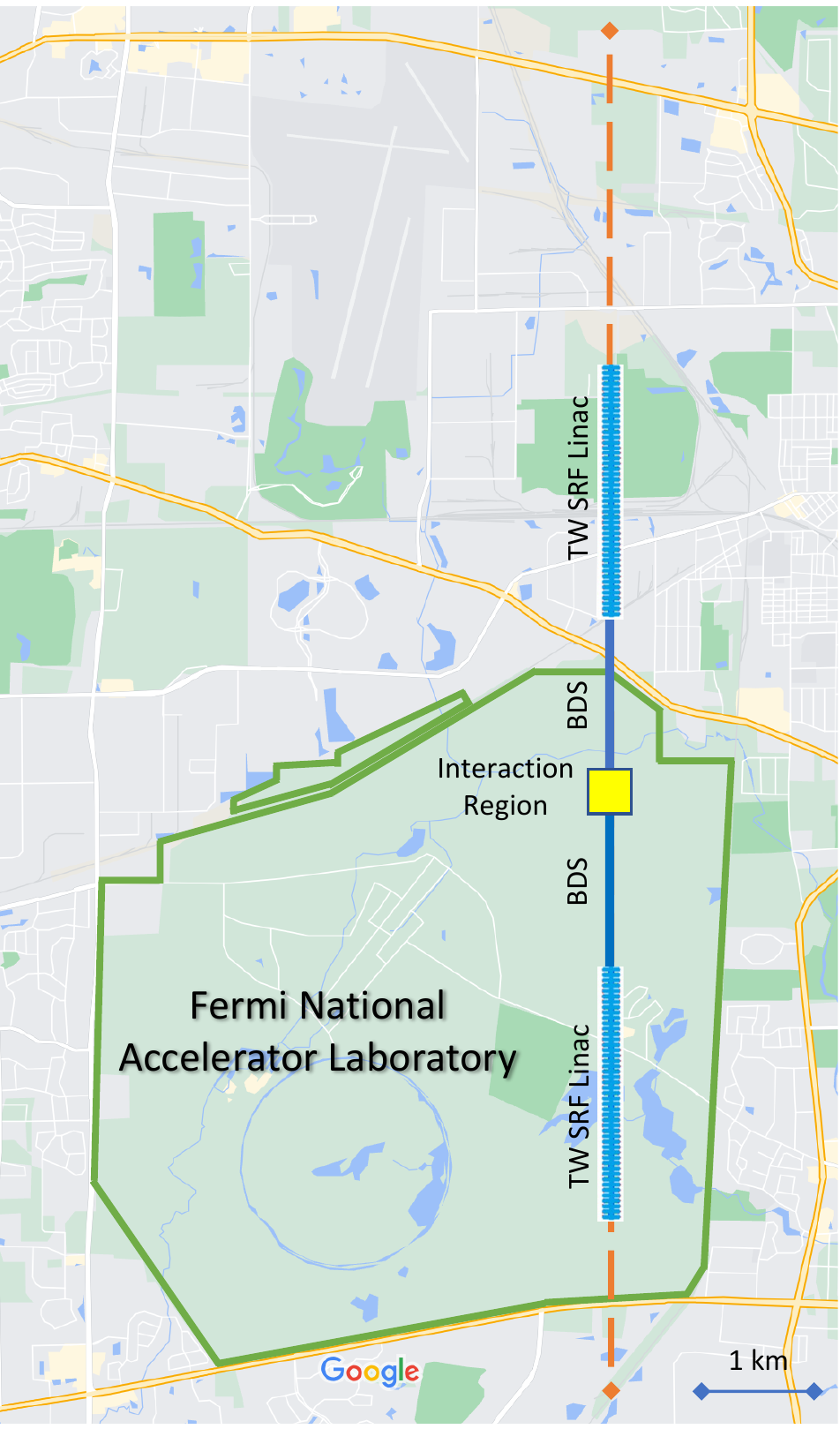}
    \caption{Possible siting of the 250~GeV HELEN collider at Fermilab. The TW option is shown. The orange dashed line indicates a 12-km stretch that might be available for a future  upgrade of HELEN to 500 GeV.}
    \label{fig:HELEN_Fermilab}
\end{figure}

Since the ILC SRF linac parameters were baselined in 2013 \cite{ILC_TDR-v3-I, ILC_TDR-v3-II}, the community made advances in further developing the technology. Three possible paths have been identified that could lead to a more compact SRF linear collider, Higgs-Energy LEptoN (HELEN) collider  \cite{HELEN}. The options are listed here in the order of their maturity:
\begin{itemize}
    \item With recent advances in surface treatments of niobium SRF cavities and development of more efficient standing wave structure geometries, it is anticipated that cavities can reach 50~-- 60~MV/m. With just 2--3 years of intensive R\&D, one can anticipate demonstration of such gradients in 9-cell SRF cavities. Assuming that cavities with operating gradient of 55~MV/m can be manufactured with sufficient yield, a 250-GeV linear collider will be 9.4-km long and will fit within the 12 km footprint in the N--S orientation at Fermilab similar to that shown in Figure~\ref{fig:HELEN_Fermilab}. The maximum energy that could potentially be reached by fully occupying 12 km is 350~GeV.
    \item New optimized traveling wave SRF structure can potentially reach accelerating gradient of $\sim$70~MV/m. We consider this option as a baseline for HELEN. At 250~GeV, the collider length is 7.5~km and it will comfortably fit within the 12-km N--S corridor as shown in Figure~\ref{fig:HELEN_Fermilab}. If we can move the IR further North, then it would be possible to upgrade the HELEN collider energy to 500~GeV while still fitting within the 12 km footprint available. %(Figure~\ref{fig:HELEN500_Fermilab}).
    \item If the 90 MV/m gradient potential for Nb$_3$Sn cavities with $Q$ of $1\cdot10^{10}$ at 4.2~K (based on extrapolations from high power pulsed measurements) can be realized, then the 250-GeV collider would fit entirely on the Fermilab site along one of NE--SW diagonals as shown in Figure~\ref{fig:LC_7kmLayout}. Alternatively, it can be built along the N--S line which offers possibility of energy upgrades. 
\end{itemize}

%\begin{figure}
%    \centering
%    \includegraphics[width=0.8\textwidth]{HELEN500_at_Fermilab.pdf}
%    \caption{The 500~GeV HELEN collider would fit into the 12-km corridor.}
%    \label{fig:HELEN500_Fermilab}
%\end{figure}

Utilizing one of the three options, one could  design and build a linear collider Higgs Factory that partially lays on the Fermilab site. R\&D program and a demonstrator test facility that would be needed to realize such a collider are described in subsequent sections and in the dedicated Snowmass2021 white paper \cite{HELEN}.

\subsection{ILC in the US}
%[Sam, PB]
Another proposal that continues to be extremely viable is the construction of ILC in the U.S. ILC has been characterized as a ``shovel-ready'' project, with well-established technical design and with world-class accelerators like the European XFEL acting as large-scale demonstrations of key SRF systems. As described above, U.S. scientists are involved in international collaborative efforts to realize the ILC in Japan. However, if ILC in Japan isn't realized, constructing ILC in the U.S. could be an attractive option. There are existing international technical coordination teams already working together from different regions across the world, discussing the next steps for ILC leading up to construction. Funding agencies are already engaged. If ILC in Japan does not proceed, enthusiasm from the U.S. HEP community could motivate funding agencies to develop plans to support construction domestically, with international contributions built on well established collaborations and frameworks. Experience from construction of the ILC-like LCLS-II accelerator involving SLAC, Fermilab, and JLab could help alleviate typical concerns of ballooning costs and schedules from projects with less well-established technologies. This includes critical expertise of and confidence in technical vendors (such as those for SRF cavities and RF power couplers) to help build confidence that cost and schedule estimates are realistic. For these reasons, the U.S. is well positioned to take on a host role for the ILC.

\begin{figure}
    \centering
    \includegraphics[width=0.9\textwidth]{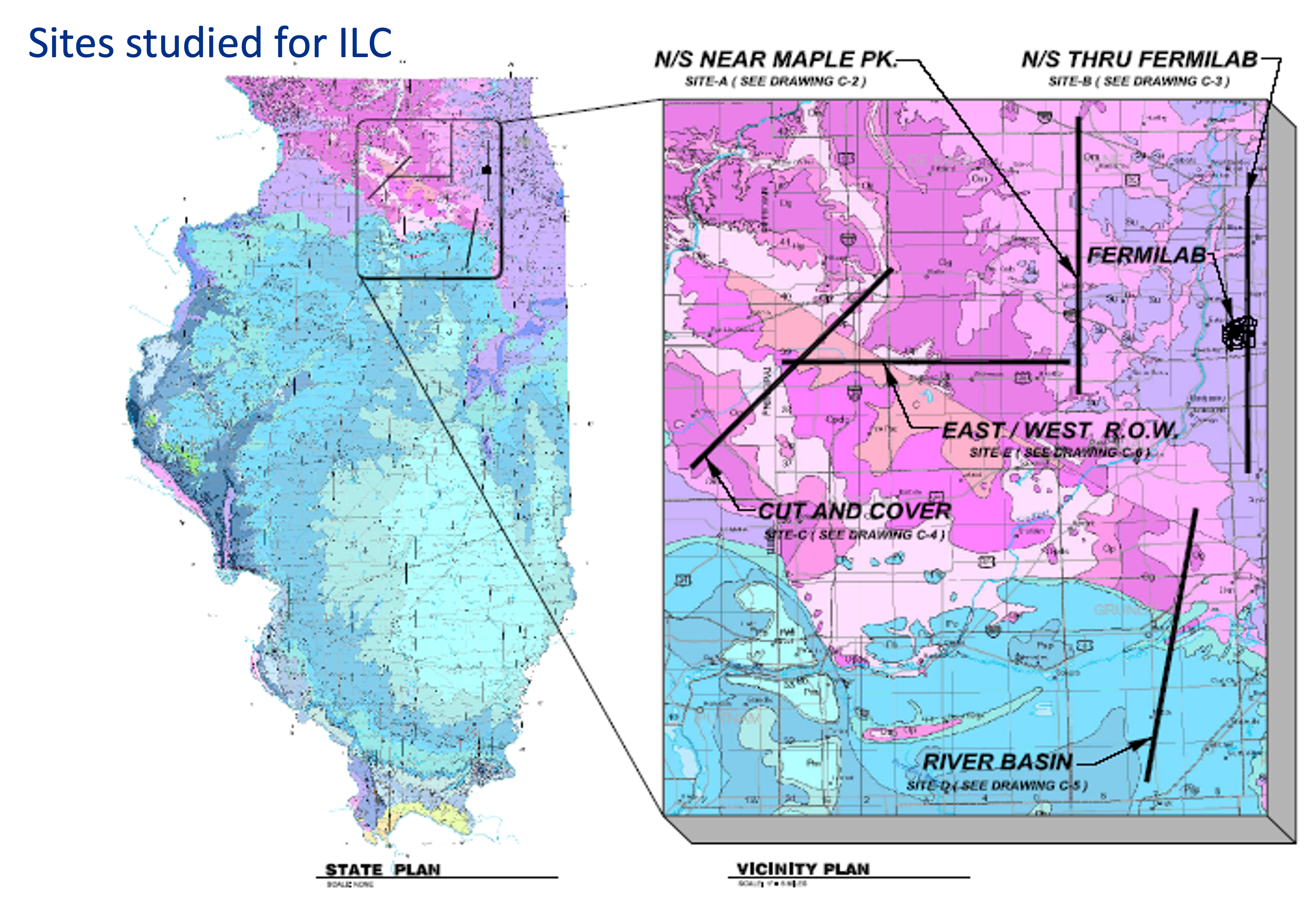}
    \caption{Potential siting options considered in the past for ILC at Fermilab.}
    \label{fig:ILC_Sites}
\end{figure}

The sites previously considered to host the ILC at Fermilab are shown in Fig.~\ref{fig:ILC_Sites}.

\subsection{Test demonstrator for  $e^+e^-$ Linear Collider}
%[Sergei N, Sergey B., Nikolay]
IOTA/FAST is an R\&D Facility for Accelerator Science and Technology at Fermilab. It has two components: an Integrable Optics Test Accelerator (IOTA), 150 MeV electron / 2.5 MeV proton storage ring \cite{IOTA}, with a dedicated proton injector and a FAST SRF linac. The 300-MeV FAST linac serves as an injector of electrons for IOTA and provides beam to dedicated experiments with linac beam. 

Beside a 8-cavity ILC-style SRF cryomodule, the electron linac includes a 5-MeV RF photoinjector of a DESY/PITZ design, a 25-m long low-energy ($\leq 50$~MeV) beam line with 2 SRF capture cavities, and a $\sim$100-m long high-energy beam line. Both beam lines are equipped with high-precision beam instrumentation.

Originally, the ILC-style FAST SRF linac was envisioned as a demonstration facility to test and operate a full ILC ``RF unit'' with ``ILC beam intensity.'' The RF unit consists of 2 cryomodules driven by a single 10~MW klystron. However, only one cryomodule was installed at FAST. The ILC beam intensity is a $\sim1$~ms long train of $\sim 3,000$ bunches (3~MHz bunch repetition frequency) with a charge of 3.2~nC per bunch. A bunch train repetition rate is 5~Hz. An r.m.s. bunch length is 300~$\mu$m.
The FAST linac was the first to demonstrate the performance of a large-scale SRF system with average beam accelerating gradient matching the ILC specification of 31.5~MV/m \cite{FASTrecord}. 

FAST can serve as a demonstrator facility for all linear colliders mentioned in this white paper. Its high-energy beam line has plenty of space to accommodate additional test cryomodules. Here is a brief explanation on how FAST can be used for technology demonstrator tests in the two linear collider scenarios:
\begin{itemize}
    \item While some upgrades to the laser and low level RF systems are needed for stable operation with full ILC bunch trains, the facility is uniquely positioned as a demonstrator for the proposed HELEN collider, which shares the beam parameters with ILC. New SRF cryomodule(s) could either replace the existing CM2 cryomodule or be added to the high-energy beam line. Additional RF system(s) will have to be installed in the latter case.
    \item \CCC~demonstrator cryomodules and associated high-power RF equipment can easily fit into the high-energy beam line tunnel. The facility has a dedicate cryogenic system, which includes a 5,800~gallons ($> 26,000$~liters) LN$_2$ tank, with a capacity exceeding the \CCC~demonstrator requirements \cite{C3demo}. At first, FAST can be used for cryogenic RF testing of the \CCC~cryomodules with and without beam. However, the present FAST injector cannot provide the \CCC~beam, and an upgrade with S-band injector would be required for a full-scale beam demonstration.
\end{itemize}

\section{An $e^+e^-$ circular collider at Fermilab}

\subsection{Design Overview}

\begin{figure}[htb]
\centering
\includegraphics[width=0.8\textwidth]{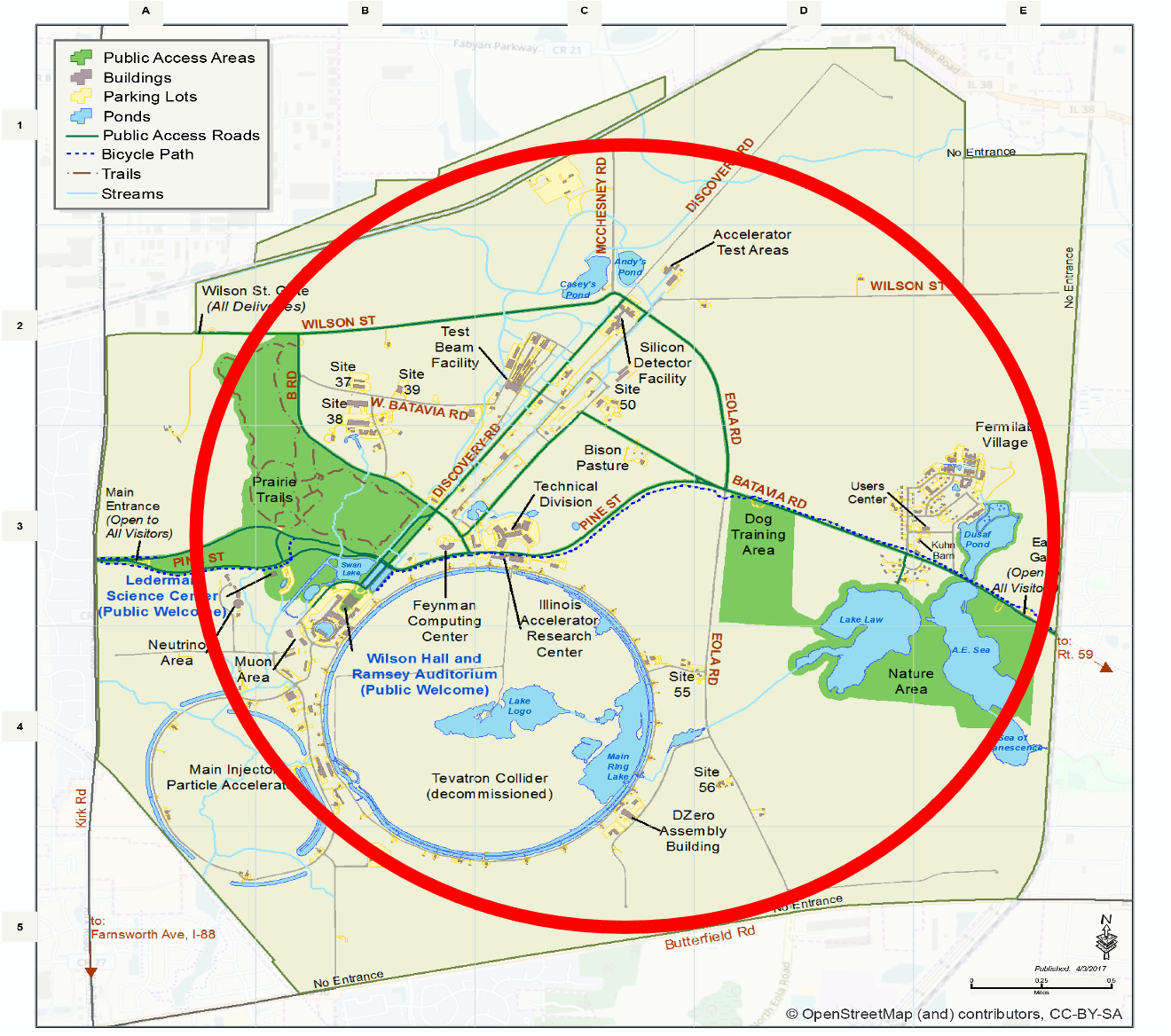}
\caption{\label{eliana-f1} Fermilab site showing the proposed 16-km site-filler collider ring.}
\end{figure}

Here we discuss the design of an $e^+e^-$ circular collider to fit within the Fermilab campus. 
Figure~\ref{eliana-f1} shows an aerial view of the laboratory site. The red circle denotes the designated location of the  proposed 16 km ring which could work as a Higgs factory at 120 GeV beam energy.
The present description is primarily based on preliminary studies presented at HF2012, and
updated in 2021. At 45.6 GeV the ring could work as a Z factory collider. 

\subsection{Design of the Higgs and Z factories}

The design principles of the Higgs factory $e^+e^-$ circular collider operating at a center of mass energy of 240 GeV is largely determined by the tolerable levels of the synchrotron radiation power, $P_T$. 
The beam current  $I$ and luminosity ${\cal L}$ in this high energy regime are given by

\begin{equation}
  I   =  \frac{e\rho}{2 C_{\gm}E^4 P_T} \\
  {\cal L}\gm^3   =  \frac{3}{16\pi r_e^2 (m_e c^2)} \;\;
\left[\rho \frac{\xi_y P_T}{\bt_y^*} H(\bt_y^*, \sg_z) \right]
\end{equation}

The luminosity equation shows that at a given energy, the luminosity is determined by the factors in square
brackets. In addition to $P_T$, these are the bend radius $\rho$, the vertical beam-beam parameter $\xi_y$,
the vertical beta function $\bt_y^*$, and the hourglass factor $H(\bt_y^*, \sg_z) \le 1$, which is a measure of the overlap between colliding bunches at the collision point. We have assumed head-on collisions between the beams which is a valid assumption with a small number of bunches in each beam. 

After fixing the maximum synchrotron radiation power to 50 MW per beam, the luminosity of the Higgs factory at Fermilab was maximized by the following choices, some of which are enforced by the limited circumference.
\begin{itemize}
    \item A single Interaction Point: In addition to a reduced cost with only one detector, this has several
  accelerator  physics advantages which include: 
\begin{itemize}
    \item a larger bending radius $\rho$ in the arc cells 
    \item total beam-beam effects (tune shift, beamsstrahlung, Bhabha scattering) are minimized;
    \item the IR chromaticity is reduced, which  will increase the momentum acceptance and consequently the beam lifetime.
\end{itemize}
    \item Very small vertical beam size at the IP (0.2 $\mu$m).
    \item Large number of particles distributed into two bunches for maximizing the luminosity. The single  bunch  intensity must respect limits set by the Transverse Mode Coupling Instability (TMCI) and by the allowable  beam-beam tune shift.
    \item Head-on collisions for operational simplicity and cost reduction.
\end{itemize}

The Z factory, which will operate at a lower center-of-mass energy of 92 GeV, is not necessarily limited by the synchrotron radiation power so it can operate at the beam-beam limit. The luminosity at this limit is given by
\begin{equation}
  {\cal L} = \frac{\pi}{r_e^2}M_B f_{rev}
\left(\frac{\kappa\bt_x^*}{(\bt_y^*)^3}\right)^{1/2} (\gm \xi_y)^2 \eps_x
\end{equation}
where $M_B$ is the number of bunches, $\kappa$ is the emittance coupling ratio. In this regime, the luminosity is proportional to the horizontal emittance $\eps_x$. This favors increasing $\eps_x$ either by lowering the phase advance per FODO cell to say 60$^{\circ}$,  or by external means such as with noise or using wigglers. This regime also requires distributing the beam current over as many possible bunches as possible which lowers the bunch intensity. To avoid parasitic collisions, a crossing angle  at the IP may be necessary and a multi-bunch feedback system may be required to avoid instabilities.

\begin{table}[htb]
\begin{center}
\begin{tabular}{|| l | c  | c || }  
  \hline
     &  Higgs Factory  & Z factory \\ \hline
Circumference [km] &16 & 16  \\
Beam energy [GeV] & 120 & 45.6  \\
Total synchrotron radiation power [MW] &  100 &  60  \\
Beam current [mA] & 5. & 140 \\
$N$ [$10^{11}$] & 8.3  & 1.67 \\
Number of bunches & 2 & 279  \\
$\beta_x^*$ [m] /  $\beta_y^*$ & 0.2 m / 1 mm  & 0.2 m / 1 mm \\
$\epsilon_x$ / $\epsilon_y$ [nm] & 21 / 0.05  & 26.1 / 0.065  \\
$\sigma_z$ [mm] & 2.9 (SR) &  6.45 \\
beam-beam tune shift per IP & 0.075/0.11 &  0.032 / 0.045 \\
RF frequency [MHz] & 650 & 650 \\
RF voltage [GV]  & 12 &  0.24 \\
Momentum acceptance  (RF) [\%] & $\pm$3  &   $\pm$9 \\
$\tau_{bs}$ [min]  & 9 - 36 &   \\
$\tau_{Bhabha}$ [min]  & 8.7 & 37  \\ 
${\cal{L}}$ per IP [$10^{34}$ cm$^{-2}$s$^{-1}$ ]&  1.0 &   6.3  \\
Production cross-section & 200 fb  & 61 nb \\
Particle production/year & Higgs: 39751  & Z: 7.64 $\times 10^{10}$   \\
\hline
\end{tabular}
\caption{Parameters of the 2012 Fermilab $e^+e^-$ Higgs and Z  Factories}\label{HF_param}
\end{center}
\end{table}

Table \ref{HF_param} shows a set of consistent parameters for both the Higgs and Z factories. The
particle production estimates assume $2 \times 10^7$ secs/year. 

\begin{itemize}
    \item The bunch length, $\sigma_z$, quoted in the table results purely from the synchrotron radiation in the arcs. This does not include the slight lengthening ($\sim 10\%$) due to beamsstrahlung and consequently the hourglass factor calculated here could be slightly  optimistic.
    \item The bunch population assumed is well below the expected beam-beam limit and the TMCI threshold. Both these limits may need to be revisited with detailed simulations. 
    \item The arc cells are 90$^0$ FODO cells, which could be replaced by the  lower emittance ones adopted in modern synchrotron radiation rings.
    \item The short beam lifetime calls for \emph{top-up injection} which ensures high average luminosity but at the cost of a full energy injector to be housed in the same tunnel.
  \end{itemize}

\subsection{Staging options}

A staged approach could envisage the use of existing machines and infrastructure as much as possible. At HF2012 some possible injection scenarios were presented. The minimal one involved the use of the Fermilab  Booster and Main Injector, in addition to a new 400 MeV linac and a positron accumulator. Besides the technical feasibility, the compatibility  with proton operation for neutrino production must be understood. The most ambitious scenarios envisaged a 1 GeV Linac, one $e^+$ accumulator ring and a superconducting RCS.

\subsection{Challenges}

\noindent Beam Dynamics: The IR nonlinear chromaticity correction system must ensure a sufficient dynamic aperture and energy acceptance. This should be achievable with only 1 IP in the ring. Next, simulations of the impact of beamsstrahlung  on the bunch parameters must be done. It is possible that the head-on crossing scheme must be changed to a \emph{crab waist} scheme, which requires the beams to cross at an angle. Its feasibility has been proven at DA$\Phi$NE and more recently at Super-KEKB. In this case in addition to  synchro-betatron resonances, simulations have found two new instabilities: a  \emph{3D flip-flop} instability in the presence of beamsstrahlung, and  a \emph{beam-beam head-tail instability}, confirmed by observation at SuperKEKB.

Other challenges include: proper positioning of rf cavities in the ring to avoid saw-tooth orbits due to energy droop,  management of the synchrotron radiation (15 kW for both beams) with a large photon critical energy (2 MeV), HOM heating in presence of large bunch population in short bunches etc.  These issues were deemed to be be manageable for the similar LEP3.

\subsection{Upgrade options}

\begin{figure}[ht]
\label{fig:lumi-circ}
\centering
    \includegraphics[width=0.65\textwidth]{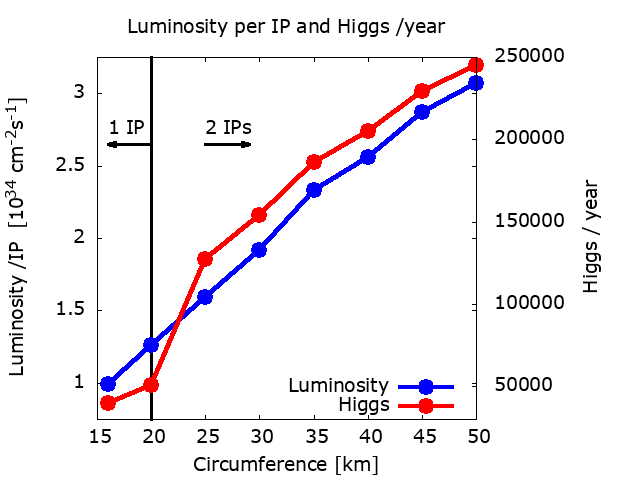}
    \caption{Luminosity per IP and the total number of Higgs per year produced from all IPs as a function of the circumference.}
\end{figure}

We consider the luminosity reach of a larger collider based at Fermilab. Figure~\ref{fig:lumi-circ} shows the luminosity per IP and the total number of Higgs produced from all IPs as the circumference increases from 16 km to 50 km. We assume that the number of IPs can be increased from 1 to 2 for circumferences greater than 20 km. Over this range, the luminosity per IP increases in the same ratio as the increase in circumference. At the larger sizes, it is possible to optimize the design for higher immensities than the values shown in this figure. As a possible future upgrade, the site-filler ring could serve as an injector for a larger collider. 

\newpage

\section{Muon collider options at Fermilab}
\subsection{Conceptual design}
\begin{figure}
\begin{center}
\includegraphics[width=0.6\textwidth]{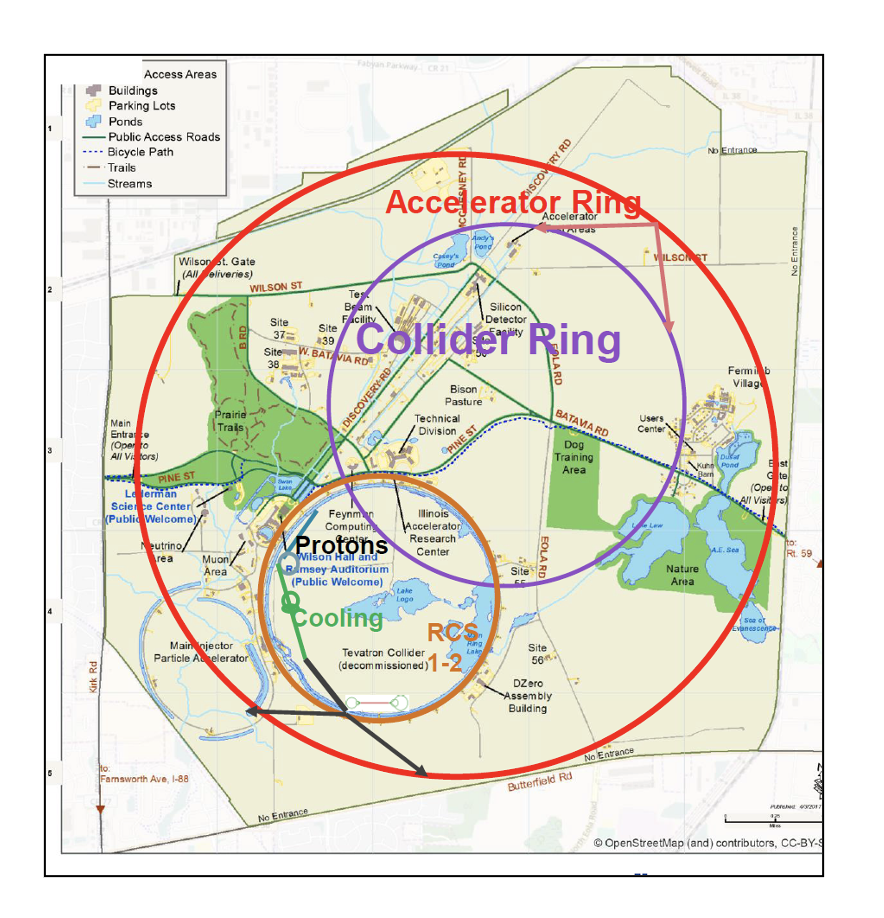}
\includegraphics[width=0.6\textwidth]{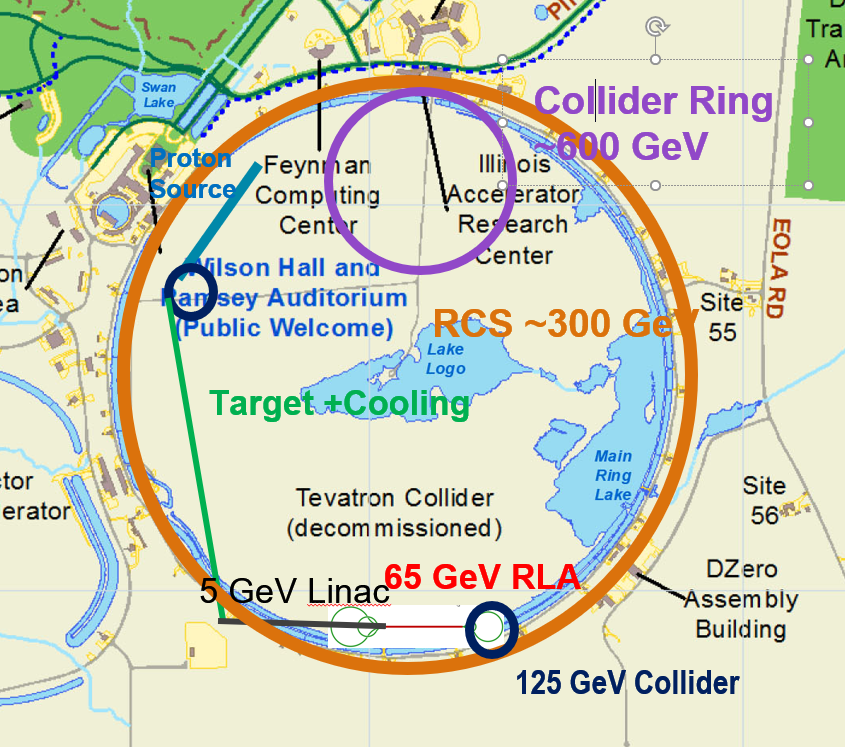}
\caption{A schematic view of the Fermilab site and the layout of the proposed collider complex for the Muon Collider site-filler (top) and a zoomed-in version showing the 125 and 600 GeV staging options (bottom).}%
\label{fig:FnalMC}
\end{center}
\end{figure}

The idea of having a MC as a potential site filler for Fermilab dates back to the early 2000’s. The focus then was a 4 TeV machine. Recently, the required parameter space towards a 10 TeV MC site filler has been identified and a first design concept has been developed. A schematic layout of this configuration is shown in Figure~\ref{fig:FnalMC}. 
The idea is to start with a future PIP-II upgrade as a proton driver. This could well align with recent proposals for a Fermilab booster upgrade~\cite{Nagaitsev:2021xzy} or extension of the PIP-II linac~\cite{Belomestnykh:2021oyc}. The target will operate at around 8 GeV with a 5 Hz repetition rate and a beam power around 2 MW, although this can be reduced if more cooling is achieved. 6D muon cooling can be achieved with a rectilinear channel first, followed by a solenoidal 4D cooling channel using NC RF at 325 MHz and 650 MHz~\cite{Wang:2015xoa}. Muon Acceleration is achieved in three stages: (1) A Linac (up to 5 GeV) first that is followed by a Recirculating Linac (up to 65~GeV). This energy would be sufficient for a Higgs Factory~\cite{Neuffer:2013wrd}. (2) A set of two Rapid Cycling Synchrotrons that can fit into the Tevatron ring tunnel and are capable of delivering an energy up to 1 TeV. (3) A final RCS that has a radius of 2.65 km (site filler) and can bring the energy up to 5 TeV. Acceleration will be conducted with superconducting rf cavities at frequencies of 650~MHz and 1300 MHz. Based on extrapolations from Ref.~\cite{IMCC} the 10 TeV collider is expected to have a radius of 1.65 km. It is important to emphasize that given the 3 accelerator stages, staging is possible and operation at 125 GeV, 600 GeV (for the top quark Yukawa measurement), and 2--3 TeV can be envisioned as intermediate states. Figure~\ref{fig:FnalMC} shows a schematic view of the collider for the different stages.

\subsection{Recent Technology Advancements}
There have been several technological accomplishments over the last decade or so. Below we highlight some of them:
\begin{itemize}
    \item \textbf{Liquid Mercury Targets:} The MERIT experiment~\cite{Kirk:2008zza} provided a proof-of-principle demonstration of a target system based on a free mercury jet inside a 15-T solenoid and showed that is capable of sustaining proton beam powers of up to 4 MW.
    \item \textbf{NC RF in 3 T Field:} The experiment conducted at Fermilab MTA facility~\cite{Bowring:2018smm} demonstrated stable high-vacuum, normal-conducting RF cavity operation at gradients of 50  MV/m in an external magnetic field of 3 T, through the use of beryllium cavity elements. A high-pressure Hydrogen gas filled RF (HPRF) cavity was also demonstrated with intense beams in a multi-Tesla solenoid field at MTA.~\cite{PhysRevLett.111.184802}. Cooling simulations show that the HPRF cavity can be used in various ionization cooling schemes~\cite{PhysRevAccelBeams.20.032002}.
    \item \textbf{Rapid Cycling Magnets:} A High Temperature Superconductor (HTS) based fast cycling prototype accelerator magnet was demonstrated to operate up to about a 300 T per second ramping rate with some 0.5 T field in the magnet gap~\cite{Piekarz:2021mna}. 
    \item \textbf{Ionization Cooling:} Demonstration of ionization cooling by the Muon Ionization Cooling Experiment (MICE) at RAL. %citation?
    \item \textbf{Lattice Design:} Self-consistent lattice designs of the various subsystems have been delivered. These included the Front-end and Cooling systems~\cite{Stratakis:2014nna}, Acceleration scenarios, ~\cite{Bogacz:2017iia} and Collider Rings up to 6 TeV~\cite{Wang:2015xoa}.  
\end{itemize}

\subsection{Future R\&D needs and Synergies}
\begin{itemize}
\item \textbf{Proton driver:} Fermilab’s PIP-II program will be capable of delivering beam powers up to 1.2 MW. Several proposals are under development for either expanding the Linac (PIP-III) or combining the existing linac with RCS so that to increase the beam power to > 2 MW. The ESS MW proton linear accelerator can be upgraded and extended to demonstrate the generation of a nanoseccond-scale beams with very high charge (10$^{15}$) proton pulses that can be used for the generation of the initial muon pulses required for a muon collider. 

\item \textbf{Target:} Fermilab has an active target development program, including targets for Mu2e-II (100 kW) and LBNF (1.2-2.4 MW).  The Mu2e-II geometry is a simpler version of the MC target system, with targets within high field large-bore solenoids.  The field strength of Mu2e-II solenoids is lower and the target length is shorter than the MC target system. However, making the Mu2e-II target system is still extremely challenging. Fermilab also hosts RaDIATE to explore targets for LBNF at 2.4 MW operation. The Fermilab research can collaborate with the Mu2e-II target group and with RaDIATE to synergetically develop the target technology for the MC. 

\item \textbf{Cooling:} %Demonstrations of the performance of RF cavities in magnetic fields are crucial. 50 MV/m at 3 T has been demonstrated at the MTA. Further tests are needed to establish performance at the parameters of cooling scenarios. Higher field magnets are needed at the cooling aborbers. 
%MAP considered cooling with magnetic fields up to 30 T. (Commercial MRI magnets are now available at 29 T and the record field demonstrated is 32 T with bores similar to those needed for cooling; these could be extended to MC parameters.)
Improving the cooling performance is a primary goal of the cooling design R\&D. Depending upon the future target system, decay, bunching, and phase rotation (called the  "front end"), the following 6D cooling channel must be optimized. Improving cooling can significantly relax the beam requirements, reducing the primary proton beam power, the beam induced background at the collider detector, and the neutrino flux. Research on integration of AI techniques can aid in making the channels shorter and perhaps identify new parameters for improved cooling. Different cooling schemes such as the Parametric resonance Ionization Cooling (PIC) scheme for cooling to ultra low emittances~\cite{PIC} or the FOFO Snake~\cite{FOFOSnake} for cooling both muons simultaneously should be explored. 

\item \textbf{Acceleration:} A RCS will require the operation of high-gradient RF cavities. While 1300~MHz SRF at 35 MV/m has been demonstrated for ILC cavities, 50 MV/m would be desired for a site filler. RCS accelerators will also require fast cycling magnets at rates of 500--1000~T/s with peak fields of up to 4 T. Fermilab has demonstrated 290 T/s but at a lower field (0.6~T). While a recirculating linac (RLA) acceleration scheme to 65 GeV has been shown, more design studies are needed to demonstrate RCS acceleration towards TeV energies. FFAs could also be developed for fast muon acceleration. 

\item \textbf{Collider Ring:} 
%The collider ring requires 16 T arc dipoles with a 15 cm bore. Moreover, neutrino flux mitigation is a concern. In addition, MAP only has studied lattices up to 6 TeV. The US-MDP program will have ID  120 mm, 12-15 T dipole demonstrators with Nb3Sn coils within the next 3-4 years. 
The TF and EF groups can investigate the physics cases at 600 GeV, 3 Tev, and 10~TeV center of mass energy. A new collider lattice must be designed. Possible solutions to mitigate neutrino flux are to make the collider at ~100 m depth, add magnets or move the lattice. %Improving cooling can significantly relax these parameters.      

\item \textbf{Magnet:}
MAP considered cooling with magnetic fields up to 30 T. (Commercial MRI magnets are now available at 29 T and the record field demonstrated is 32 T with bores similar to those needed for cooling; these could be extended to MC parameters.)
The collider ring requires 16 T arc dipoles with a 15 cm bore. Moreover, neutrino flux mitigation is a concern. In addition, MAP only has studied lattices up to 6 TeV. The US-MDP program will have ID  120 mm, 12--15 T dipole demonstrators with Nb3Sn coils within the next 3--4 years.
\item \textbf{RF cavity:}
Demonstrations of the performance of RF cavities in magnetic fields are crucial. 50 MV/m at 3 T has been demonstrated at the MTA. A LN$_2$ temperature cavity will have a potential to reach high RF gradient in stronger magnetic fields than the past demonstration. Further tests are needed to establish performance at the parameters of cooling scenarios. Integrating RF cavities with cooling magnets is a crucial engineering challenge. 
High power RF sources need to be developed. 

\end{itemize}

\subsection{Higgs Factory Considerations}
A muon collider Higgs factory continues to be of interest to the community, especially if none of the $e^{+}e^{-}$ options are realized. Such a machine can substantially improve precision of most Higgs boson couplings when compared to HL-LHC. It can also be complementary to $e^{+}e^{-}$ by providing very precise and model independent measurements of the Higgs boson total width, mass, and the muon Yukawa coupling. There is considerable overlap between the accelerator complex required for a 125~GeV Higgs factory with that required for a multi-TeV machine. Based on MAP studies, the proton driver, the front-end, and the 6D muon cooling system can be shared with a Collider. As a result, a Higgs Factory can serve as an acceleration demonstrator for subsequent higher energy stages. Moreover, acceleration will be based on more established methods, such as the use of RLA’s, and the Collider Ring circumference will be only $\sim300$~m. The final 6D cooling system, which trades off increased longitudinal emittance to obtain smaller transverse emittance  is required for a TeV-scale MC,is not needed for the Higgs factory.

\subsection{Fermilab site option for demonstrator}
\label{sec:MCdemo}
\begin{figure}[h]
\begin{center}
\includegraphics[width=0.6\textwidth]{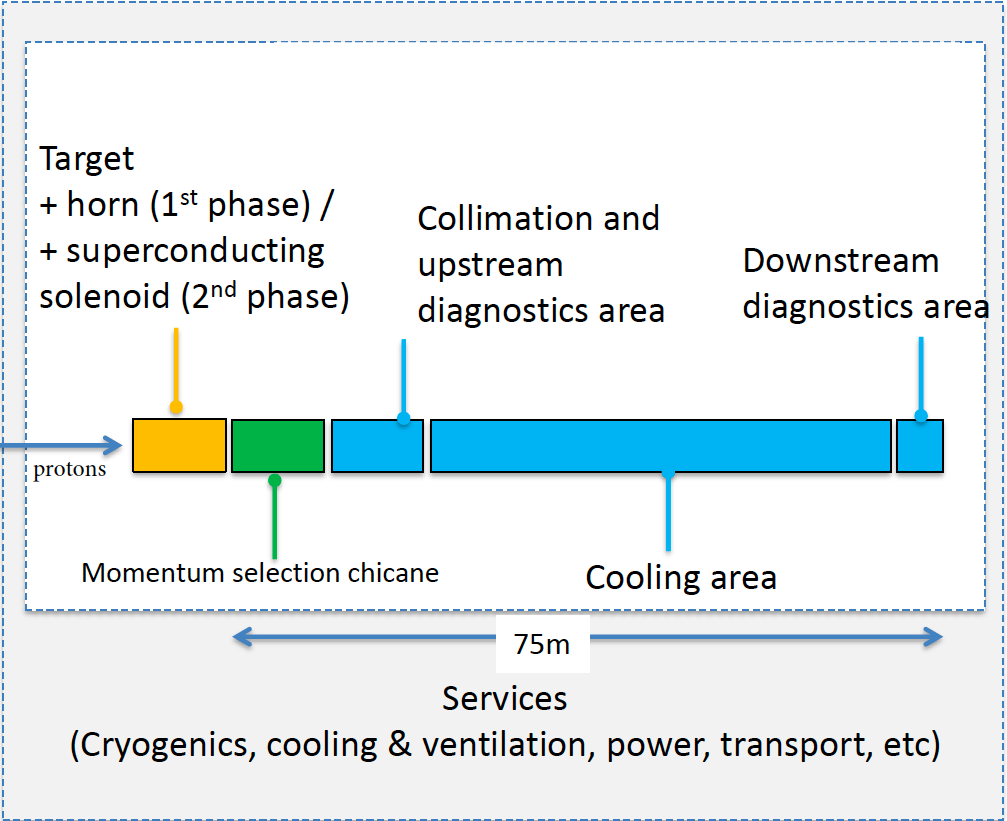}
\caption{Components of a demonstrator facility for the Muon Collider}
\label{fig:Demo_comp}
\end{center}
\end{figure}

\begin{figure}[h]
\begin{center}
\includegraphics[width=0.5\textwidth]{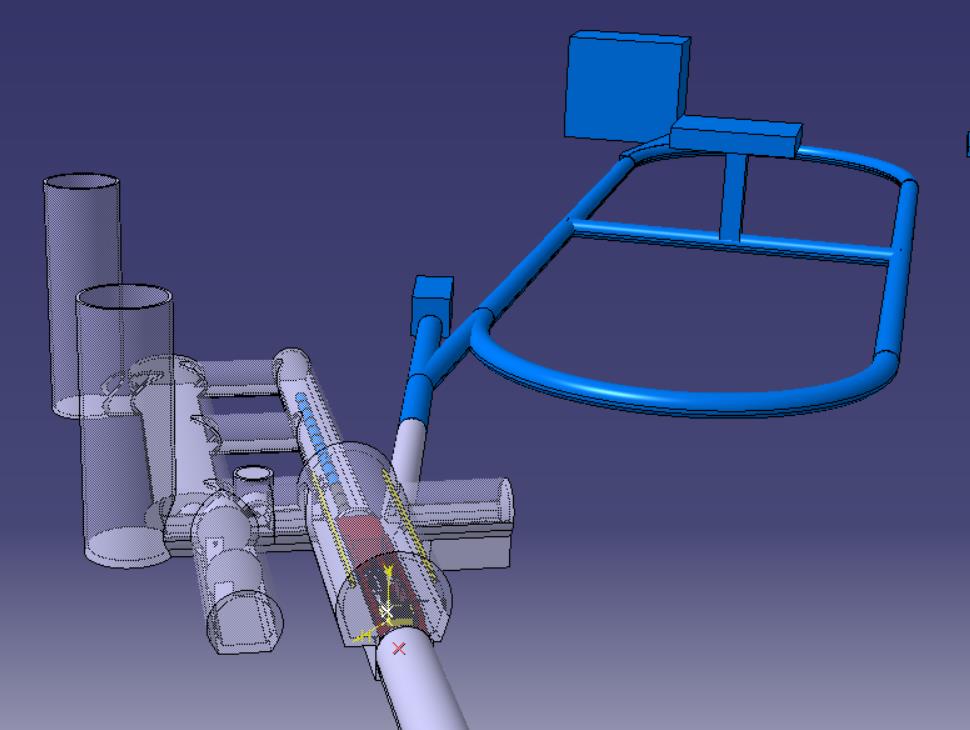}
\caption{Schematic of the demo facility feeding muons to nuSTORM.}
\label{fig:Demo_nustorm}
\end{center}
\end{figure}

A fundamental component of the R\&D for a Muon Collider is a late-stage 6D cooling demonstrator.  This was true during MAP and now is a central component of the IMCC.  Within the IMCC, a great deal of work has been done to define what a demonstrator facility should be. The IMCC is taking a modular approach to the facility where initially a minimum configuration is deployed and over time upgrades are implemented that deliver additional capability.  The demonstrator facility components as defined by the IMCC are indicated in Figure~\ref{fig:Demo_comp}
The facility includes a $\simeq$~100~kW target stations (upgradable to higher power), a pion momentum selection section, collimation and a demonstration cooling section.  The facility will be designed with flexibility in mind so that different cooling lattices can be tested. Given the envisioned configuration of the facility, it could support HEP experiments. Branching off from the target station, the facility could support nuSTORM~\cite{Rogers:2021wyv} and/or ENUBET\cite{Brizzolari:2022seb}.  Figure~\ref{fig:Demo_nustorm} shows how a demo facility could be used to feed nuSTORM.

\begin{figure}
\begin{center}
\includegraphics[width=0.99\textwidth]{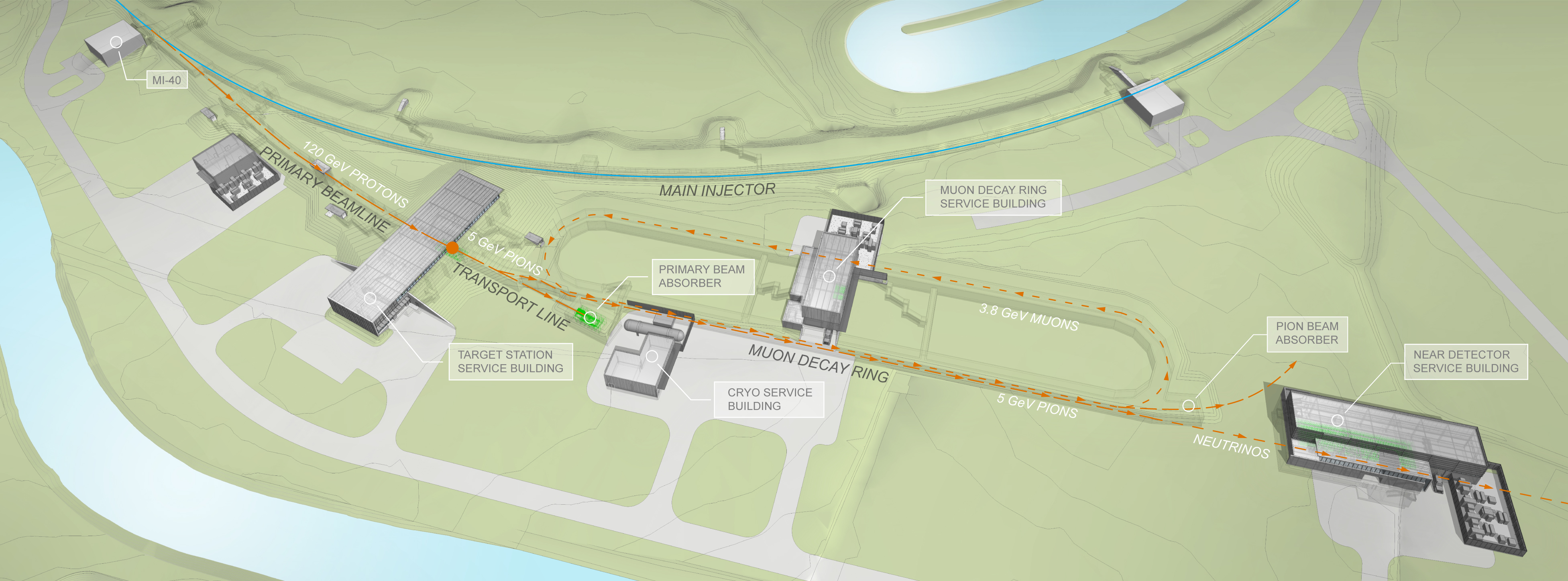}
\caption{nuSTORM at Fermilab facility components.}
\label{fig:nuSTORM_F}
\end{center}
\end{figure}

The IMCC design assumes siting at CERN where protons are extracted from the PS using land close to the TTf10 line.  However, the nuSTORM siting plan at Fermilab~\cite{Adey:2017dvr} using the Main Injector offers many attractive features for supporting a Muon Collider demo facility feeding an experimental neutrino program with a beam from muons or kaons. Figure~\ref{fig:nuSTORM_F} shows the Fermilab nuSTORM facility layout that was fully detailed in the nuSTORM Project Definition Report~\cite{Lackowski:2013ria}.  It is easy to see how a muon cooling demo facility could be fed by the nuSTORM target station.
\newpage

\section{A proton-proton collider  at Fermilab}

We consider here the possibility of building a pp collider to fit on the Fermilab campus to operate at energies about twice that of the LHC . The Tevatron and its injector complex can serve as the entire injector chain for this collider. Given the compact circumference of 16 km, this will require dipole fields of unprecedented
strength. Simply scaling from the LHC  circumference and dipole field strength shows that dipole fields around 28 T would be required to reach  energies close to 28~TeV in the Fermi site filler. This is far beyond the scale of fields envisaged in the design of other future pp colliders such as the FCC-hh and the SppC. Nevertheless, we will proceed with the bold (and likely foolhardy) assumption that such magnets can be built with the required accelerator quality and in a cost effective and timely manner. With the major issue swept under the rug, we discuss the accelerator physics of this collider.

\subsection{Design of the pp collider}

The design of the arc lattice requires, among other choices, selecting the cell length and dipole length. A longer dipole generally leads to lower magnetic fields but is limited from above to $\sim 15$~m for logistical reasons. We chose a dipole length of 12 m and a cell length of 76 m which result in dipole fields at the lower end of a range. The design of the interaction region (IR) is more complex and will be done when necessary. The parameters of this collider discussed below are obtained without an IR design. 

Due to the large number of bunches required to attain  high luminosities in this collider, crossing angles need to be introduced at the interaction points to avoid parasitic collisions. Assuming a crossing angle $\theta_c$ in the horizontal plane, the luminosity ${\cal  L}$ and beam-beam tune shifts $(\xi_x, \xi_y)$ are given by
\beqr
    {\cal L} & = & \fr{f_{rev} n_{b} N_p^2}{4 \pi \sg_x^* \sg_y^*} R(\theta_c)  \\
    \xi_x  & = &  \fr{r_p N_p \bt_x^*  R(\theta_c)^2 }{2 \pi \gm \sg_x^*(\sg_x^* + R(\theta_c)\sg_y^*)} ,
    \;\;\;     \xi_y =   \fr{r_p N_p \bt_y^*  R(\theta_c)^2 }{2 \pi \gm \sg_y^*(\sg_x^* + R(\theta_c)\sg_y^*)}
    \\
    R(\theta_c) &  = & \fr{1}{\sqrt{1 + (\theta_c \sg_z/(2 \sg_x^*))^2 } }
\eeqr
Here $n_b$ is the number of bunches, $N_p$ is the bunch intensity, $\sg_x^*, \sg_y^*$ are the rms transverse sizes at the IP, $R(\theta_c) \le 1$ is the reduction factor due to the crossing angle and $\sg_z$ is the rms bunch length.  The bunch intensity decreases  during a luminosity store with the loss rate given by
    \beqr
    \fr{d}{dt}N_p = - n_{IP} \sigma_{tot}^{pp} \fr{{\cal L}}{n_b}
    \eeqr
Here $n_{IP}$ is the number of IPs and $\Sigma_{tot}^{pp}$ is the total pp cross-section. At the high energies of this collider, synchrotron radiation has a dominant effect on the beam dynamics as is discussed below. The emittance damping is modeled simply as an exponential decay $ \eps_{\perp}(t) = \eps_0 \exp[-t/\tau]$ where $\tau$ is the emittance damping time and $\eps_0$ is the initial emittance. 

    \vspace{2em}
    
    \noindent \underline{Design Assumptions}
    \bit
    \item The arc lattice is based on FODO cells, 90$^{\circ}$ phase advance per cell.
    \item Two insertions for experiments, with a total length of 2.6 km for all the straight sections.
    \item The beam separation at the long-range interactions in the drift space before the first IR quadrupole is
    12$\sg$, larger than the separation of 9.5$\sg$ in the LHC.
    \item The maximum beam-beam tune shift in any plane from all IPs is 0.025, based on Tevatron experience.
    \item The crossing angle is in the horizontal plane at one IP and in the vertical plane at the other IP.
    \eit

\begin{table}[htb]
\begin{center}
\begin{tabular}{|| l | c  | c | c | c || }  
  \hline
     &   $E_{CM} =24$ TeV &  $E_{CM} =27$ TeV & HE-LHC & FCC-hh  \\ \hline
Circumference [km] &16 & 16 & 27   & 97.8   \\
Beam energy [TeV] & 12  & 13.5 &  27 & 50 \\
Number of IPs &  2 &  2 & 2 & 4  \\
Main dipole field [T] &  24.4    & 27.4 & 16 & 16   \\
Number of bunches & 1600  & 1600   & 2808 & 10600  \\
Harmonic number &   21348  &  21348  & 35640 &   \\
Bunch spacing [ns] &  25   &  25   & 25 &   25  \\
rms  emittance $\eps_{\perp}$ [mm-mrad] & 1.5  & 1.5 & 1.38 & 2.2  \\
rms bunch length $\sigma_z$ [cm] & 3.7  &  3.6 & 7.55   & 8  \\
$\beta_x^*, \beta_y^*$ [m]  & 0.5, 0.5   &  0.5, 0.5 & 0.45 0.45 &  1.1, 1.1 \\
Beam current [mA] & 446 &  333   & 1120 &  500   \\
Particles/bunch $N$ [$10^{11}$] &  0.93   &  0.69  & 2.2 & 1.0  \\
Beam energy [GJ]  &  0.29   &  0.24  & 2.4 & \\  %    & 8.4  \\
Crossing angle [$\mu$rad] &  184     &   173  & 185 &  104  \\
Initial b-b tune shifts/IP $(\xi_x, \xi_y)$ &  (0.0066, 0.0072) &  (0.005, 0.0054)  &  & 0.0055  \\
Max. b-b tune shift from 2 IPs &    0.024   & 0.025   &     &   \\
Trans.  emittance damping time [hrs] & 1.8  & 1.3  & 1.1  &  \\
Critical energy of synch.  rad. [keV]   &  0.377 &    0.537   &  & \\
Synch. rad. power/ beam [MW] &  0.043    &   0.0   &  2.4  & \\
Density of synch. rad in arc [W/m] &  4.2    &  5.1 & 3.74   & 28.4  \\
Initial  ${\cal{L}}$/IP [$10^{34}$ cm$^{-2}$s$^{-1}$ ] & 3.2 &  2.0  & 15 &  5. \\
Peak  ${\cal{L}}$/IP [$10^{34}$ cm$^{-2}$s$^{-1}$ ] &  3.5   & 2.85   &   &  \\
Number of events/crossing &  80  &  50  & 800 &  170   \\
Initial beam lifetime from burn-off [h] &  6.4    &  7.6    & 3.0 &  \\
Debris power into IR magnets [kW] &  6.2     & 4.4     &   & \\  
\hline
\end{tabular}
\caption{A set of parameters each for the pp collider at 24 and 27 TeV in the Fermi site filler compared to the baseline
  parameters of the FCC-hh collider.}
\label{pp_param}
\end{center}
\end{table}

Table \ref{pp_param} shows the parameters at two center of mass energies of 24 TeV and 
27 TeV  and compared to the FCC-hh collider. 
\bit
    \item The transverse emittance damping time is on the scale of an hour. This damping time $\sim 1$~hr is much less than  the emittance growth due to intra-beam scattering and will have some  beneficial effects. The small beam size   will not require cooling  and should also help against instabilities.
    \item Synchrotron radiation power at 44 kW is an order of magnitude larger than in the LHC but two orders of magnitude less than in the FCC-hh. Consequently, the problem of removing the synchrotron radiation will  be challenging but perhaps manageable.
    \item The critical energy of synchrotron radiation is also about an order of magnitude larger than the critical energy of 43 eV in the LHC. This will significantly impact the production of electrons by photo-absorption at the beam pipe and other surfaces. Electron cloud generation and associated instabilities will need significant mitigation efforts. Nevertheless, this problem will be less severe than in the FCC-hh.
    \item Debris power into the IR magnets is $\sim$ 4--6 times the value in the LHC. This should be manageable with improvements in the design of absorbers and machine protection systems in the IR.
    \item The  number of interactions per crossing increase $\sim$ 2--3 fold from the 32 events in the LHC, but  is much less than in the FCC-hh.
\eit 
\bfig[htb]
\centering
\includegraphics[scale=0.65]{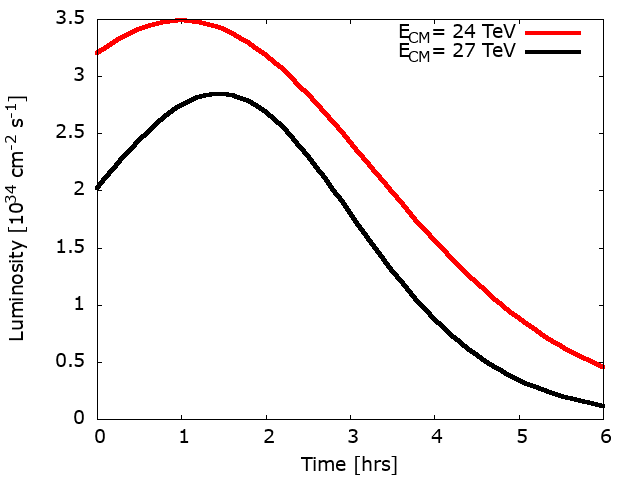}
\includegraphics[scale=0.65]{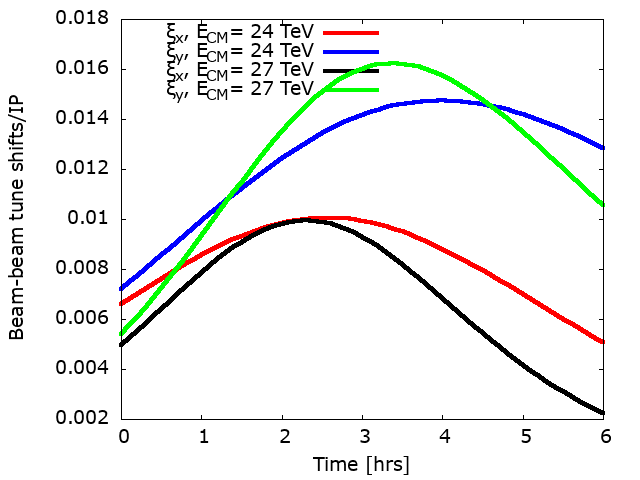}
\caption{Evolution of the luminosity (top) and the beam-beam tune shifts with a crossing angle in the
  horizontal plane (bottom) at center of mass    energies 24 TeV and 27 TeV.}
\label{fig: lumi}
\efig
\clearpage

Figure \ref{fig: lumi} shows the evolution of the luminosity and the beam-beam tune shifts over 
a 6 hr store. The time dependence arises both from  particle losses from burn off and the emittance decay from radiation damping. The luminosity increases for about 2 hrs before decreasing to about 10\% of the initial luminosity after 6 hrs. This plot suggests that each store time should not exceed $\sim 4$~hours. The emittance reduction has a stronger impact on the beam-beam tune shifts;  e.g.  at 27~TeV $\xi_x$ increases by a factor of 2 while $\xi_y$ increases by nearly a factor of 4. This large increase in the beam-beam tune shift poses a major limit on the achievable luminosity. The emittance change in these calculations ignores emittance growth mechanisms such as intra-beam scattering which has a growth time  $\sim 6$ hrs, thus the increase in beam-beam tune shift is somewhat exaggerated.

Beam-beam compensation   with electron lenses would be effective in reducing the head-on tune shift and increasing the luminosity.

\subsection{Challenges}

Clearly the largest challenge is to design and build dipole magnets with fields at and  above 24 T together with the required field quality. The next major challenge is to keep the cost of the collider, with all components, to be within reasonable limits. All other issues are relatively insignificant compared to these two.  

    \vspace{2em}
    
\noindent \underline{Accelerator Physics Challenges}:
\bit
\item  Machine protection: Very high beam energy and magnetic energy, improved \& sophisticated collimation,
\item Novel diagnostics for halo control and beam loss, monitor radiation damage, photon absorbers to protect
  cold magnets and equipment
\item  High synchrotron radiation: Impact on components, cryogenic system, radiation hard electronics, electron
  cloud
\item Beam dynamics: electron cloud effects, compensation of beam-beam interactions (head-on and
  long-range), instabilities during injection and the ramp, dynamic aperture, ...
\eit 

\subsection{Upgrade options}

Compensation of the head-on beam-beam tune shift with electron lenses would increase the luminosity, as
mentioned above. Crab cavities would restore head-on collisions and also raise luminosity. After a few years of
operation, it should be possible to increase luminosity by standard methods such as lowering $\bt^*$,
reducing the crossing angle etc. Finally, this collider can serve as an injector to a collider operating at the 100 TeV
energy scale.

\section{Technology R\&D Directions}
\label{SEC:TRD}
\subsection{Introduction}
As the requirements for colliders continue to grow, the need for investment in accelerator and detector technology research and development becomes more critical. ``Brute force'' approaches to colliders -- by making extremely large rings or long linear tunnels -- are possible, but only feasible up to a point. Investments in R\&D can pay off multiple fold. For example, developing stronger superconducting magnets would benefit not only hadron colliders, but a muon collider as well; or high gradient / high $Q$ SRF cavities will find applications across several fields from HEP to nuclear physics, to FELs, to industrial accelerators. R\&D time frames are difficult to predict, and in same cases, there are large advances that can be leveraged quickly. A recent example was the development of nitrogen doping for SRF cavities, bringing an increase in quality factors by a factor of $\sim3$. This was a crucial factor for the feasibility of the LCLS-II accelerator, which began production using nitrogen doping less than 5 years after its invention. In this section we describe some promising directions and give approximate time frames, with the caveat that time frames have both positive and negative error bars.

\subsection{Magnet R\&D}
%[GA]

The circular nature of some of the colliders under consideration, such as muon  colliders and high energy proton colliders (FCC-hh or SppC) naturally drives the focus to the study and development of advanced magnets in various configurations (dipoles and quadrupoles, solenoids, fast ramping magnets, etc) and at high field levels, normally enabled by the use of superconducting technology. In addition, the number of magnets -- in some cases highly specialized, one-of-a-kind elements, in others several hundreds or thousands of cost-efficient and reproducible magnets -- drive considerations on the best way to produce such magnets for the machines described in this white paper.

Superconducting magnets (dipoles and quadrupoles) based on Nb$_3$Sn technology have been demonstrated up to $\sim15$~T (single units). Hybrid solenoids using NbTi, Nb$_3$Sn and high-temperature superconductor (HTS) tape technology have been demonstrated up to 32~T.
All the magnets mentioned above are produced in National Laboratories in single quantities or in “boutique” operations in quantities of few dozens in the 2020’s, such as for the Nb$_3$Sn focusing magnet production for the HiLumi Project at the LHC.

A muon collider based on fast ramping magnet for muon acceleration would require the magnets shown in Table \ref{magnets_tab1}. A high energy hadron collider would require the magnets shown in Table \ref{magnets_tab2}.

\begin{table}[h]
\caption{Approximate fields and quantities of magnets for a Muon Collider}
\label{magnets_tab1}
\begin{center}
\begin{tabular}{ |l|c|c| } 
 \hline
 Magnet type & Field & Quantity \\ 
 \hline
 Production target EHF solenoid	 & $\sim 40$ T	& 1 \\
Cooling channel VHF solenoids &	$\sim 25$ T	& Dozens\\
Cooling channel HF solenoids	& $\sim 4-15$ T	& Hundreds\\
Fast ramping magnets	& $\Delta B \sim 2$ T and d{B}/dt $\sim 1000$~T/s	& Few Hundreds\\
MR high field dipoles	& $\sim 8-12$ T	& Few Hundreds\\
IR high field quadrupole & $\sim 15-16$ T	& Dozens \\
 \hline
 \end{tabular}
\end{center}
\end{table}

\begin{table}[h]
\caption{Approximate fields and quantities of magnets for a very high energy hadron collider}
\label{magnets_tab2}
\begin{center}
\begin{tabular}{ |l|c|c| } 
 \hline
 Magnet type & Field & Quantity \\ 
 \hline
MR high field dipoles & $\sim 15-16$ T	& Few Thousands\\
IR high field quadrupoles	& $\sim 15-16$ T	& Dozens \\
 \hline
\end{tabular}
\end{center}
\end{table}

In the muon collider, individual solenoids at very high magnetic field  (32 T and above) may not necessarily need to be superconducting in nature and partially resistive (albeit power-hungry) solutions can be considered for those close to the strong radiation environment of the production target. On the other hand, the cooling channel and all the remaining magnets in the muon collider and very high Energy hadron collider have to rely on superconducting technologies. The above considerations are exposing the two main challenges in addressing the feasibility of such future colliders in the next decade.

\underline{Industrialization challenge:}
When needed quantities are in the “hundreds/thousands of units”, industrialization is a must to maintain the necessary cost control and insure uniformity of deliverables. Such aspect was already identified as a challenge for Magnet R\&D in the 2014 P5 report. This challenge applies to several beam-line magnetic elements listed above (Main Ring dipoles, fast-ramping magnets, cooling solenoids, etc) and the approach has to involve laboratories and universities in the R\&D and prototyping phases, but needs to demonstrate a feasible technology transfer and an appropriate industrialization process for the pre-series and series production phases.

\underline{Field level challenge:}
When a high or very high magnetic field level is necessary to ensure the technical success of machine elements and yet the number of units is small (focusing IR magnets, few dozens of very high field solenoids, etc.) an approach based on laboratories or universities involvement from R\&D to final production can be entertained given the inherent difficulties in technology transfer of high field magnets applications.

\subsubsection{HTS, LTS/HTS magnets}

R\&D efforts on superconducting magnets have been energized, especially in Europe, following the 2019 update of the European Strategy for Particle Physics and its identification of FCC-ee, FCC-hh, and muon colliders as viable venues of exploration for future machines.

In the U.S., the GARD\footnote{GARD is the General Accelerator R\&D program sponsored by the U.S. DOE Office of HEP.}-supported nation-wide Magnet Development Program (MDP) together with the Conductor Procurement and R\&D (CPRD) are pursuing generic R\&D with 4 primary goals to explore the performance limits of Nb$_3$Sn accelerator magnets, develop and demonstrate an HTS accelerator magnet with a self-field of 5~T or greater (to use in an hybrid configuration with a Nb$_3$Sn magnet), investigate fundamental aspects of magnet design and technology, and pursue Nb$_3$Sn and HTS conductor R\&D.

At FNAL, the mentioned above generic MDP efforts are materializing in a series of specific thrusts with the following elements related to future muon or hadron colliders:

\begin{itemize}
\item Stress-managed cos-theta (SMCT) coils developed for Nb$_3$Sn and Bi2212 16+~T magnets \cite{COMB}
\item  20 T hybrid design studies for LTS\footnote{Low Temperature Superconductor}-HTS magnets and Development of new technology (HTS, REBCO SC based, COMB) for 18+~T hybrid magnets
\item Development of HTS-based fast-ramping magnets \cite{CORC}, \cite{STAR}
\item Development of  Nb$_3$Sn APC (artificial pinning centers) wires with higher stability and critical current $J_c$ at or above FCC specs \cite{Nb3SnFCC} and development of high-$C_p$ wires with good drawability.
\item Research on coil assembly materials, such as thermoplastics and high-toughness resins
\item Development of fiber optics technology as cryogenics strain gauges and temperature sensors.
\item Development of a capacitor-based device (QCD) to improve training behavior in Nb$_3$Sn superconducting magnets and usage of AI \cite{AI} to detect the quench precursors and  other state of the art magnet diagnostics tools
\end{itemize}

The previous and other generic magnet R\&D efforts are described in a white paper submitted by the MDP Collaboration \cite{MDP_WP}.

\subsubsection{Fast-ramping magnet R\&D}
%[inputs from H. Piekarz]
Next generation HEP facilities such as muon colliders, future circular colliders, and high-intensity proton synchrotrons for neutrino research demand substantially faster cycles of beam acceleration then available at present.  To date, the highest ramping rates achieved in the operational superconducting accelerator magnets based on the LTS (NbTi) are about 4 T/s, a limitation caused by a very narrow allowable operational temperature margin.

Fast-ramping HTS-based magnets offer a cost-effective solutions for many future particle   accelerators mentioned above but especially for the acceleration of the short-lived particles such as muons. The AC losses in the fast-ramping accelerator magnet are due to power losses in both the magnet energizing conductor and the magnetic core. The power losses in the magnetic core can be reduced by using as thin as practically possible laminations. The power losses in the conductor can be reduced by minimizing both its mass and exposure area to the ramping magnetic field descending from the core. Application of the superconductor allows to strongly minimize magnet cable mass and size, and as a result also the size and mass of the magnetic core. Very importantly, however, the HTS conductor can be set to to operate at 5 K, well below its critical temperature of e.g., 30 K, providing in this way a wide operational temperature margin and facilitating the temperature-based quench detection and protection systems. 
%The conceptual design of the HTS-based accelerator magnet is shown in Fig. 1. The vertically arranged 2-bore magnet is powered with a single conductor coil. 
A prototype HTS-based accelerator magnet of 0.5 m length and two beam gaps of 100 mm (hor.) x 10 mm (vert.) was successfully tested \cite{FRMPiekarz}.  Preparations are now underway to increase this test magnet $B$-field to 0.9 T and the ramping rates up to (500--600) T/s.

Future goals in the next 2 years include upgrade the present HTS test magnet to 2~T or higher $B$-field and the $dB/dt$ rates up to 500--1000 T/s. In the longer term (3--6 years), goals should include the design, construction and power test a long prototype magnet as required for the muon accelerator and the initiation of a  possible industrialization process \cite{FRMWhite}.

%Reference to Fast Ramping Magnet Whitepaper ??

\subsubsection{LEAF Program}

In order to transition from the generic R\&D effort described above to meet the industrialization and field level challenges described in the introduction to this section, an effort based on the magnets leading edge technology, yet driven to demonstrate the feasibility of future colliders, is essential. 

Historically, the development and demonstration of maturity of the Nb$_3$Sn technology for application in HL-LHC was made possible by a 15 years-long (2003--2018) DOE investment in an U.S. national program of directed R\&D (called LARP) working in combination with generic and complementary R\&D efforts (called, at that time, CDP, GARD, etc.). 

In the same spirit, the proposed Leading-Edge technology And Feasibility-directed (LEAF) program is foreseen to be a decade-long effort to be concluded on the time-scale of ~2034--2035.  The LEAF Program describes the hand-off from the generic magnet R\&D effort to a feasibility-directed approach entertaining a more directed design and development effort and, where necessary, a downselection and industrialization effort for large quantities production. The LEAF Program is described in a separate white paper submitted to this Snowmass process \cite{LEAF}. The main elements of the LEAF program can be summarized as following:

\begin{itemize}
\item Design and developments of magnets addressing specific elements for the colliders under consideration (field and field quality, aperture, operations, radiation environment, interfaces with experiments, etc.)
\item Support for industrial production and usage of advanced superconductor (LTS and HTS) 
\item Scaling of magnet lengths (fast-ramping magnets, SMCT, MR and IR magnets, ...)
\item Synergetic collaboration for high field solenoid developments with other offices in DOE or with NSF
\item Industrialization and cost reduction through next generation design for Nb$_3$Sn magnets \cite{2ndGEN}
\end{itemize}

\subsection{RF R\&D}
%[Sam, Sergey B]
Advanced RF systems are central to a large number of proposals for future HEP facilities. This includes future colliders like ILC, FCC-ee, CEPC, CLIC, C$^3$, HELEN, FCC-hh, SppC, and muon colliders, as well as drivers for intensity frontier experiments like LBNF/DUNE. It also includes some smaller-scale experiments such as axion haloscopes. The needs for RF R\&D aren't only in the area of increasing gradient -- other important areas to improve include cavity quality factors,  RF source power efficiency and cost, and RF control systems. Mitigating issues related to short- and long-range wakefield effects is important, especially for high-intensity machines. 

A decade-long roadmap for RF R\&D was developed under the framework of the DOE GARD program in 2017 \cite{GARD-RF-Strategy}. The roadmap was worked out by a team of leading researchers in the field from various national labs and universities, both domestic and international. The roadmap reflects the most promising research directions for advances that enable future experimental high energy physics programs. While much progress has been made since that time, most of the topics remain valid. However, the roadmap should be updated and extended into the next decade according to the needs of future HEP machines.

In this section, we divide R\&D topics into SRF cavities, normal conducting RF (NCRF) cavities, and companion topics.

\subsubsection{SRF for future colliders}
SRF cavities are used to accelerate beams in some of the most advanced worldwide accelerator facilities, including for HEP (such as the LHC and PIP-II), basic energy sciences (European XFEL, LCLS-II, SNS, ESS) and nuclear physics (CEBAF, FRIB, EIC). SRF R\&D over the years has led to performance improvements that have enabled new applications which previously had not been feasible. Continued investment in SRF R\&D can help to increase the scientific reach of colliders in different ways.

Increasing accelerating gradients, while maintaining high quality factors, is a key R\&D direction. Higher gradients allow linear accelerator tunnels to be shorter and use fewer components to reach a given energy. This helps to enable both linear colliders (e.g., ILC and its upgrades, HELEN) and pulsed drivers for machines like muon colliders and intensity frontier experiments. Promising R\&D directions are being pursued for increasing gradient, including new superconducting materials, travelling wave cavities, new cell shapes for standing wave structures, cleanroom robotics to reduce field emission, layered superconductor structures, and new impurity doping treatments, as well as more fundamental explorations of the limits of RF superconductivity, such as the use of ``slow'' surface materials that could prevent dissipation from magnetic flux penetration (for examples of SRF R\&D directions, see the many Snowmass LOIs and white papers on the subject in the Accelerator Frontier, such as \cite{SRFRDSnowmass2021_01,SRFRDSnowmass2021_02,SRFRDSnowmass2021_03,SRFRDSnowmass2021_04,SRFRDSnowmass2021_05,SRFRDSnowmass2021_06,SRFRDSnowmass2021_07,SRFRDSnowmass2021_08,SRFRDSnowmass2021_09,SRFRDSnowmass2021_10,SRFRDSnowmass2021_11,SRFRDSnowmass2021_12,SRFRDSnowmass2021_13,SRFRDSnowmass2021_14}). These efforts are funding-limited -- there are many exciting ideas to pursue and not enough resources.

Increasing quality factors of SRF cavities is another key R\&D direction. Higher quality factors reduce RF dissipation to the liquid helium. This can reduce the cryogenic plant size (which can have a substantial impact for continuous wave RF accelerators like FCC-ee and CEPC), or allow pulsed accelerators to operate with higher duty factor. Promising R\&D directions that are being pursued include new superconducting materials, new impurity doping treatments, and expulsion of magnetic flux to minimize trapped flux dissipation.

A very important issue for high-intensity machines (e.g., FCC-ee and CEPC) is to mitigate effects of higher-order mode (HOM) impedances of SRF cavities on stability of beam motion. Developing HOM-damped SRF cavities (sometimes called single-mode cavities) and components to couple out and absorb HOM power is an important R\&D topic for these machines, see e.g, \cite{Belomestnykh2012}.

\subsubsection{NCRF for future colliders}

The main challenge in NCRF for future linear colliders is developing high-gradient structures with an acceptable breakdown rate and adequate mitigation of wakefield effects. The CLIC team has developed and demonstrated a room-temperature X-band structure stably operating at $\sim~70$~MV/m. Further improvements in gradient has been possible by cooling down copper structures to cryogenic temperatures, which strengthens the material and improves the breakdown rate. \CCC~follows this path with developing novel C-band structures \cite{nanni2021c}. However, there are still many R\&D issues to address, which are described in the Snowmass white paper \cite{CCCdemonstrator}.

NCRF for a muon collider faces a very specific challenge of operating high-gradient cavities in high magnetic field of the muon cooling channel. Some R\&D has been done in the past, but more is required to find an optimal combination of the cavity frequency, geometry, material, operating temperature and pressure.

\subsubsection{Companion R\&D topics}
RF cavities require RF power sources, for which two areas of R\&D can be beneficial: cost and efficiency. Improving RF power source efficiency can be especially beneficial for accelerators that have high AC power requirements, which may be dominated by the RF system demand.

As gradients increase, it is important to perform R\&D on corresponding improvements in auxiliary systems that will need to be modified in order to take full advantage of the higher gradients. These include high-power RF distribution, resonance control systems, mitigation methods for field emission, and RF power couplers.

\subsection{High Power Targetry R\&D}
%[Katsuya, TBD ]
A High-Power Target (HPT) system is a critical beam element to accomplish future High Energy Physics experiments. 
Future neutrino facilities, like LBNF and J-PARC, propose 1--3 MW proton beams delivered to a neutrino target \cite{lbnf2016,jparc2017}. 
The beam power range is comparable to a muon collider and neutrino factory, which propose 2--5 MW proton beams \cite{mcJP}. 
On the other hand, the European Particle Physics community suggests investigating a 100 TeV center-of-mass energy hadron collider FCC-hh \cite{fcccdr}. 
The HPT technology R\&D is also beneficial to the FCC-hh which requires a radiation hardened beam elements: beam collimator, beam damper, beam window, and beam instrumentation that will need to tolerate a radiation does equivalent to a MW of beam power.  Even though the FCC-hh does not have a target system in the complex, HPT R\&D is needed.
The current HEP target technology tolerates a beam power up to 1 MW. 
The goal of the proposed R\&D extends their capability beyond 1 MW beams. 

\subsubsection{Material science R\&D}
To maximize the yield of secondary and tertiary particles coming from a target system, the typical length of the target is a few interaction lengths. A hot spot appears in every beam cycle at a depth of one interaction length in the target. Such a high cycle thermal stress and radiation damage make the target lifetime short.  The RaDIATE collaboration was formed to research a radiation tolerant material for HEP solid targets~\cite{Frederique2021,Kavin2021}. 
The Post Irradiation Experiment (PIE) and Displacement Per Atom (DPA) cross-section experiment are proposed at Fermilab, BNL, and CERN to extend the fundamental radiological material science in HEP energy regimes. Graphite is currently the most popular material for a neutrino target. 
It restores a mechanical strain because it can be annealed at high temperature caused by the energy deposition of the beam. 

State of the art technology in nano-science is capable of investigating radiation damage at the atomic scale.
A recent study suggests that a compound material, such as Ti-6Al-4V~\cite{Ishida2021} or a high-entropy compound~\cite{Atwani2021} have radiation resistance by controlling the crystal phase change and irradiation temperature. 
A nano-fiber target is another possible technology to mitigate propagating thermal shock~\cite{Sujit2021}. 
Another possible solution is the use of liquid or granular materials which potentially mitigate the instantaneous thermal stress issue. 

The Muon Accelerator Program (MAP) investigated a mercury jet target. 
The concept was experimentally demonstrated at instantaneous power up to 8 MW. However, because mercury is harmful to the environment, and since the SNS and J-PARC report cavitation damage in a mercury target vessel, mercury targets are not favored. A flowing granular Tungsten pion production target is proposed to avoid the issues of a mercury target. 
A small particle Tungsten powder is injected into a beam interaction volume by using a He gas jet. 
A fluidized powder target introduces new challenges, however.  These include: achieving reliable circulation and continuous stable horizontal dense phase flow, managing heat dissipation, mitigating radiation damage and erosion of the containing pipework and beam windows, as well as ensuring reliable diagnostics and controls for the powder handling processes.

\subsubsection{Develop precise simulation tool for HEP target design}
Producing a precise hadronic interaction model in simulation is crucial for designing a target system and reducing systematic uncertainty in experiments. To this end, the experimental data (from NA61 and EMPHATIC) will be used to optimize simulation code (GEANT4, MARS and FLUKA). 
Present target design is typically a monolithic shape made by stacking either a small piece of identical rod or block. 
An optimal HEP target could have a varied cross section and material property along the target length to have better mechanical strength and secondary/tertiary yields. Artificial Intelligence (AI) can be applied to optimize the design of target systems. Utilizing a national High Performance Computing (HPC) facility supported by DOE is likely needed to optain the high statistics needed for such simulation studies. 

\subsubsection{Pion capture channel R\&D}
The pion capture channel should be addressed in the target system R\&D. The target is immersed in a solenoidal magnet to adapt captured pions at the target to the downstream beam line. 
The field strength is adiabatically reduced along the beam path length to induce a beam focusing. 
A peak field strength at the target is 15--20 T and the strength is down to 2 T in 10--20 meters at the end of the pion capture channel. A high pion yield and high capture efficiencies in this scheme are successfully demonstrated in simulation. 
To mitigate the radiation issue, the solenoid coil in a high radiation area is a hybrid structure: An inner coil is normal conducting and an outer one is super-conducting, and a thick radiation shielding layer is inserted between the two coils. However, there is no engineering design to remove the heat from the channel in a short time. 
It is unknown how long these solenoid coils, especially an electric insulator in the coil, can survive in such extreme environments. 
Besides, there is no practical design for a primary proton beam dump. 
An engineering study and demonstration test are needed. 

A magnetic horn focusing channel is considered as an alternate option. 
It is widely used for a neutrino target system. 
It has been demonstrated with a 900 kW beam operation. 
Technology is matured and can be extended to accept multi-MW beam power. An idea of making a FODO cell by combining multiple horns is considered to capture and focus both charged particles. 
The present design goals are to validate the concept and to improve pion yield and capture efficiencies in the horn scheme.

\subsection{Detectors for future colliders R\&D}
%[Petra,Zoltan ]
Detector R\&D needs for future colliders have been studied and summarized very recently in 2019 by the DOE Basic Research Needs For High Energy Physics Detector Research and Development report~\cite{BRN_Detectors} as well as by the 2021 ECFA Detector Research and Development Roadmap~\cite{ECFA_Detectors}. Main findings from these two articles have been summarized here.

\subsubsection{Tracking}
The main workhorse for Inner Tracking Systems are silicon detectors. The most important R\&D directions in this area are to achieve full integration of sensing and microelectronics, e.g. in monolithic pixelated CMOS sensors; the development of 4D capabilities for picosecond timing; radiation hardness to extreme fluences of up to $5\times10^{18}$~n\_{eq}/cm$^2$, including exploration of alternative materials; and the development of 3D-interconnect technologies; ultra-low mass support structures and cooling systems, going hand-in-hand with low-power and optical/wireless readout capabilities. To scale up to ever larger systems, especially for silicon-based calorimeters, R\&D is needed into large wafer sizes and new, lower cost materials, such as graphene or GaAs. Testing infrastructure, such as irradiation and testbeam facilities that can reach the relevant energies and fluences, are a crucial ingredients to the success of this ambitious R\&D program. Close collaboration with industry partners is becoming more and more important in order to capitalize from ongoing advances in telecommunication and to keep the cost from becoming prohibitive.

\subsubsection{Calorimetry}
We need to develop radiation-hard calorimeters with enhanced electromagnetic energy and timing resolution. We need to develop high-granularity calorimeters with multi-dimensional readout for optimized use of particle flow methods. We need to develop calorimeters for extreme environments, such as radiation, data rates and pile-up. For silicon-based calorimeters we need to reduce the passive space by developing larger wafers, smaller guardrings, and suitable mechanical structures. We need to increase the signal by designing thicker sensors with active gain, which is especially important for electron and muon colliders. We need to invest in new technologies, such as CMOS-based sensors and digital SiPMs, as well as new materials, such as GaAs. To enable very large area detectors, new advances in interconnects need to be made, such as anisotropic conductive films or PCBs made of new materials with the same CTE as silicon. Larger scale industrialization for these detectors will be needed, in particular for hadron colliders. The challenges for calorimeters based on liquid noble gases lie in developing high readout granularity for pileup mitigation and particle-flow reconstruction, picosecond timing information, and the minimization of passive material in front of the calorimeter: hundreds of tons of calorimeter need to be supporting by low-mass cryostats. For calorimeters with light-based readout the R\&D challenges are related to the development of novel Silicon Photomultipliers (SiPMs) with large spectral sensitivity and high-bandwidth semiconductors for higher radiation tolerance, as well as digital SiPMs. The development of novel crystal and liquid scintillator technologies are crucial. 

\subsubsection{Gaseous Detectors}
The main detector types are GEM, Micromegas, $\mu$-RWELL, RPC and RICH. We need to improve time and spatial resolution with long-term stability. We need to achieve tracking in gaseous detectors with dE/dx and dN/dx capabilities in large volumes with very low material budget and different readout schemes. We need to develop environmentally friendly gaseous detectors for very large areas with high-rate capability. They need to be radiation-hard. For Inner Tracking applications the detectors need to be ultra-lightweight. Given the large areas needed, the cost needs to be driven down, perhaps through industrialization. These detectors can be used for Muon Systems, inner tracking Detectors, including particle identification (PID), as well as Calorimeters and Pre-shower Detectors.

\subsubsection{Photon Detection and PID}
Here we need to develop photosensors for extreme radiation environments, in particular at hadron colliders. The leading technology for this are SiPMs, for which we have to develop low noise, fast-timing capable and inherently radiation-hard versions. R\&D should go into developing RICH and imaging detectors with low mass and high-resolution timing capabilities in order to enable particle ID. Develop compact high-performance time-of-flight detectors for particle ID.

\subsubsection{Electronics and Data Processing}
We need to develop new technologies to deal with greatly increased data density, such as high data rate ASICs and systems, and new link technologies, such as optical fibers, wireless, wireline, and free-space optics to communicate between detector layers for increased on-detector data reduction. Power consumption and readout efficiency also need to be improved. We need to continue to develop new technologies to increase the intelligence on the detector itself. This involves front-end programmability, configurability and modularity; intelligent power management and advanced data reduction techniques using AI/ML. Readout technologies need to be on par with new developments in 4D and 5D detector techniques. For example, high-performance sampling ADCs and TDCs, as well as high-precision timing distribution need to be developed. All of this needs to be developed to work in extreme radiation environments, especially for future hadron and muon colliders. Especially in the area of readout electronics and data processing, commercial developments are advancing at a record pace. HEP needs to be able to keep up with these developments to profit from industry standards and cheaper processes. 

\subsubsection{Collider Detector R\&D at Fermilab}
The Fermilab Detector R\&D program currently supports a wide range of R\&D topics in the area of collider physics. One main research focus is on the development of silicon sensors and ASICs with special interest in picosecond timing and 3D-integration. R\&D is also being performed on extruded, molded and 3D-printed scintillator with special emphasis on light-yield and radiation hardness. We are working on thermally improved carbon fiber composites for light-weight support structures. One area of our R\&D is focused on radiation-hard and B-field-hard DC-DC converters. In the area of new materials we are performing long-term ``Blue Sky'' R\&D involving GaAs and Graphene. GaAs with In quantum dots is a potential new material for photon-collecting ultra-light tracking or calorimetry detectors. Graphene, or other large-bandgap materials, have the potential to replace silicon for large-area, low-mass, cost-effective tracking detectors. Furthermore, we are developing novel readout links based on silicon photonics, and we are working towards intelligent, self-calibrating detectors using AI/ML.  

Picosecond Timing R\&D is one of the current two high-priority directions of the Fermilab R\&D program. This is being approached by a combination of sensor R\&D, ASIC R\&D, Systems engineering and facility development. On the sensor side we are working on different LGAD designs as well as the principle of small pixels that could potentially deliver 5D information (position, timing and direction). Future R\&D plans include an expanded picosecond timing R\&D program as well as increased R\&D for on-detector AI/ML. Long-term Blue Sky R\&D efforts will continue. 

Two fundamentally important components in Fermilab’s successful collider detector program are the Fermilab Test Beam Facility (FTBF) and the Irradiation Test Area (ITA). It is crucial that these facilities will be supported, maintained and improved also in the future. We are submitting a proposal for new test beam and high-intensity irradiation facilities at Fermilab as part of a Snowmass white paper~\cite{IFFacilities}. These will be designed to enable detector R\&D for future colliders.

\subsection{Software and Computing Infrastructure\label{sec:softcomp}}
%[Daniel Elvira]

Given the fast evolution of the computing hardware landscape and potential breakthroughs expected by the time future colliders are operational, defining computing models for future colliders today is way too early. On the other hand, computing needs for physics studies and accelerator and detector R\&D in the next few years are easier to predict and need to be addressed without delay. Teams with expertise on accelerator and detector simulation modeling tools focused on future colliders need to be strengthened and provided with resources within high-energy physics laboratories and university groups. Software infrastructure commensurable with the requirements to run compute intensive simulations based on beam and detector modeling tool-kits must be developed and effort spent to incorporate the necessary features for user friendly interfaces and accurate predictions.

Simulation tools must be able to model accelerator components and beam transport conditions unique to each of the proposed collider accelerators. They are of fundamental importance in the design and optimization of these components, as well as the actual configurations of R\&D experiments performed to address technology challenges. For example, in the case of the muon collider, extensive simulation would be needed to improve target and cooling channel designs and analyze the data of the associated demonstrator experiments. Event generators must be capable to model hard collisions and processes potentially occurring at the energies future colliders would be operated. Detector simulation tookits, such as Geant4,  must be improved to be able to model the complex geometries of future detectors and the physics interactions inside the detectors. Reconstruction algorithms should be developed to extract all the physics information made available by novel detector technologies and features. Even if the computing demands of future colliders were smaller than those of HL-LHC, software tools must be adapted to support prospect studies and R\&D activity. Additionally, as the computing landscape evolves, software needs to be adapted or re-engineered to incorporate modern techniques, such as Artificial Intelligence, and run on new computing hardware and platforms, including super-computing facilities requiring efficient use of hardware accelerators.

A long-term commitment to build expertise through new hires and training is not a small issue, given that the utilization of the above-mentioned computing infrastructure and the execution of the software development projects require skills and expertise which are scarce and in high demand. Continuity and predictability are essential to build competent and productive teams to provide software and computing support for future collider efforts where the US plans to play a leading role.
\section{Summary and Conclusions}
There is significant interest in the U.S. HEP community to make progress towards the construction of a global collider, to pursue precision Higgs physics and to search for new physics beyond the standard model.   There are several proposed candidates which have been extensively studied globally and they are in various stages of readiness. In addition to engaging in colliders proposed to be hosted abroad, there is great interest to explore options to host a collider in the U.S. following the LBNF/DUNE project completion. 

Of all the candidates on the table for an $e^+e^-$ Higgs factory, the ILC is the most mature and “shovel ready” project for construction. If the ILC does not get approval to move forward in Japan within this year, the ILC could be considered to be built in the U.S., at Fermilab. In this paper, we have also discussed a few other novel, timely, cost effective, compact Higgs factory options that are suitable for Fermilab site. The linear $e^+e^-$  collider options discussed are highly promising.  We also discuss a staged Muon Collider from a 125 GeV Higgs to multi TeV energy range. We present also preliminary studies for a compact site-filler Hadron Collider. 

We have discussed critical technology R\&D and demonstrator projects for \CCC~linear collider and the muon collider.  To progress towards a decision by next Snowmass on the selection of one of the collider options for the U.S., we call for an integrated collider R\&D program. The R\&D program should adequately support required design studies and focused R\&D  on the most promising collider option(s) to address major challenges in order to demonstrate feasibility in five to six years time.

\section{Acknowledgements}
Work supported by the Fermi National Accelerator Laboratory, managed and operated by Fermi Research Alliance, LLC under Contract No. DE-AC02-07CH11359 with the U.S. Department of Energy. The U.S. Government retains and the publisher, by accepting the article for publication, acknowledges that the U.S. Government retains a non-exclusive, paid-up, irrevocable, world-wide license to publish or reproduce the published form of this manuscript, or allow others to do so, for U.S. Government purposes. We appreciate contributions in the study of collider siting options by the Fermilab Facilities Engineering Services Section (FESS). 
%\end{linenumbers}

\bibliographystyle{JHEP}
\bibliography{references}

\providecommand{\href}[2]{#2}\begingroup\raggedright\begin{thebibliography}{10}

\bibitem{ILC_TDR-v3-I}
C.~Adolphsen, M.~Barone, B.~Barish, K.~Buesser, P.~Burrows, J.~Carwardine
  et~al., \emph{The {I}nternational {L}inear {C}ollider {T}echnical {D}esign
  {R}eport - {V}olume 3.{I}: {A}ccelerator {R\&D} in the {T}echnical {D}esign
  {P}hase},  \href{https://arxiv.org/abs/1306.6353}{{\ttfamily 1306.6353}}.

\bibitem{ILC_TDR-v3-II}
C.~Adolphsen, M.~Barone, B.~Barish, K.~Buesser, P.~Burrows, J.~Carwardine
  et~al., \emph{The {I}nternational {L}inear {C}ollider {T}echnical {D}esign
  {R}eport - {V}olume 3.{II}: Accelerator {B}aseline {D}esign},
  \href{https://arxiv.org/abs/1306.6328}{{\ttfamily 1306.6328}}.

\bibitem{BhatTaylorNaturePhys}
P.C.~Bhat and G.N.~Taylor, \emph{Particle physics at accelerators in the united
  states and asia}, {\emph{Nature Physics} {\bfseries 16} (2020) 380}.

\bibitem{ATLAS:2012yve}
{\scshape ATLAS} collaboration, \emph{{Observation of a new particle in the
  search for the Standard Model Higgs boson with the ATLAS detector at the
  LHC}}, \href{https://doi.org/10.1016/j.physletb.2012.08.020}{\emph{Phys.
  Lett. B} {\bfseries 716} (2012) 1}
  [\href{https://arxiv.org/abs/1207.7214}{{\ttfamily 1207.7214}}].

\bibitem{CMS:2012qbp}
{\scshape CMS} collaboration, \emph{{Observation of a New Boson at a Mass of
  125 GeV with the CMS Experiment at the LHC}},
  \href{https://doi.org/10.1016/j.physletb.2012.08.021}{\emph{Phys. Lett. B}
  {\bfseries 716} (2012) 30} [\href{https://arxiv.org/abs/1207.7235}{{\ttfamily
  1207.7235}}].

\bibitem{shiltsev2021modern}
V.~Shiltsev and F.~Zimmermann, \emph{Modern and future colliders},
  {\emph{Reviews of Modern Physics} {\bfseries 93} (2021) 015006}.

\bibitem{aihara2019international}
H.~Aihara, J.~Bagger, P.~Bambade, B.~Barish, T.~Behnke, A.~Bellerive et~al.,
  \emph{The {I}nternational {L}inear {C}ollider. {A} {G}lobal {P}roject},
  \href{https://arxiv.org/abs/1901.09829}{{\ttfamily 1901.09829}}.

\bibitem{bambade2019international}
P.~Bambade, T.~Barklow, T.~Behnke, M.~Berggren, J.~Brau, P.~Burrows et~al.,
  \emph{The {I}nternational {L}inear {C}ollider: {A} {G}lobal {P}roject},
  \href{https://arxiv.org/abs/1903.01629}{{\ttfamily 1903.01629}}.

\bibitem{ILCSnowmass2021}
\emph{The {I}nternational {L}inear {C}ollider: {R}eport to {S}nowmass 2021 (to
  be submitted)},  2022.

\bibitem{padamsee2019impact}
H.~Padamsee, A.~Grassellino, S.~Belomestnykh and S.~Posen, \emph{Impact of high
  {Q} on {ILC250} upgrade for record luminosities and path toward {ILC380}},
  \href{https://arxiv.org/abs/1910.01276}{{\ttfamily 1910.01276}}.

\bibitem{padamsee2021ilc}
H.~Padamsee, \emph{{ILC} upgrades to 3 {TeV}},
  \href{https://arxiv.org/abs/2108.11904}{{\ttfamily 2108.11904}}.

\bibitem{TESLAcavity}
B.~Aune et~al., \emph{Superconducting {TESLA} cavities}, {\emph{Physical
  Reviews Accelerators and Beams} {\bfseries 3} (2000) 092001}.

\bibitem{FASTrecord}
D.~Broemmelsiek, B.~Chase, D.~Edstrom, E.~Harms, J.~Leibfritz, S.~Nagaitsev
  et~al., \emph{{Record high-gradient SRF beam acceleration at Fermilab}},
  \href{https://doi.org/10.1088/1367-2630/aaec57}{\emph{New Journal of Physics}
  {\bfseries 20} (2018) 113018}.

\bibitem{SRF_2021}
Y.~Yamamoto et~al., \emph{{Stable beam operation at 33~MV/m in STF-2
  cryomodules at KEK}},  in \emph{{Proceedings of SRF2021}}, ({East Lansing,
  MI, USA}), 2021.

\bibitem{evans2017international}
L.~Evans and S.~Michizono, \emph{The international linear collider machine
  staging report 2017},  \href{https://arxiv.org/abs/1711.00568}{{\ttfamily
  1711.00568}}.

\bibitem{1st_TESLA_ws}
\emph{Proceedings of the 1st {TESLA} ({TeV} {S}uperconducting {L}inear
  {A}ccelerator) {W}orkshop}, 1990.

\bibitem{ICFA_IDT}
\emph{{ICFA ILC I}nternational {D}evelopment {T}eam},  2020.

\bibitem{fcc2019fcc}
A.~Abada et~al., \emph{{FCC}-ee: The {L}epton {C}ollider}, {\emph{The European
  Physical Journal Special Topics} {\bfseries 228} (2019) 261}.

\bibitem{FCC-hh}
A.~Abada et~al., \emph{{FCC}-hh: The {H}adron {C}ollider}, {\emph{The European
  Physical Journal Special Topics} {\bfseries 228} (2019) 755}.

\bibitem{Chao:2014pea}
A.W.~Chao and W.~Chou, eds., \emph{{Reviews of accelerator science and
  technology}: {Vol. 7: Colliders}}, World Scientific, Hackensack (2014),
  \href{https://doi.org/10.1142/9474}{10.1142/9474}.

\bibitem{Neuffer:1983jr}
D.~Neuffer, \emph{{Principles and Applications of Muon Cooling}},
  \href{https://doi.org/10.2172/1156195}{\emph{Part. Accel.} {\bfseries 14}
  (1983) 75}.

\bibitem{Tinlot:1965ab}
J.~Tinlot and D.R.~Green, \emph{{A Storage Ring for 10 BeV Muons}},
  \href{https://doi.org/10.1109/TNS.1965.4323677}{\emph{IEEE Trans. Nucl. Sci.}
  {\bfseries 12} (1965) 470}.

\bibitem{Ankenbrandt:1999cta}
C.M.~Ankenbrandt et~al., \emph{{Status of muon collider research and
  development and future plans}},
  \href{https://doi.org/10.1103/PhysRevSTAB.2.081001}{\emph{Phys. Rev. ST
  Accel. Beams} {\bfseries 2} (1999) 081001}
  [\href{https://arxiv.org/abs/physics/9901022}{{\ttfamily physics/9901022}}].

\bibitem{Zisman:2000dn}
M.S.~Zisman, \emph{{Neutrino factory and muon collider collaboration: R\&D
  program}}, \href{https://doi.org/10.1016/S0168-9002(01)01318-3}{\emph{Nucl.
  Instrum. Meth. A} {\bfseries 472} (2000) 611}.

\bibitem{Palmer:2013/07/02bta}
{\scshape Neutrino Factory, Muon Collider} collaboration, \emph{{An Overview of
  the US Muon Accelerator Program}},  in \emph{{International Workshop on Beam
  Cooling and Related Topics}}, 6, 2013.

\bibitem{Palmer:2016gws}
M.~Palmer, \emph{{Fast cooling, muon acceleration and the prospect of muon
  colliders}},  in \emph{{Challenges And Goals For Accelerators In The XXI
  Century}}, (Singapore), pp.~781--798, World Scientific (2016),
  \href{https://doi.org/10.1142/9789814436403_0041}{DOI}.

\bibitem{McDonald:2010zzc}
K.T.~McDonald et~al., \emph{{The MERIT High-Power Target Experiment at the CERN
  PS}}, {\emph{Conf. Proc. C} {\bfseries 100523} (2010) 3527}.

\bibitem{Bowring:2018smm}
D.~Bowring et~al., \emph{{Operation of normal-conducting RF cavities in
  multi-tesla magnetic fields for muon ionization cooling: a feasibility
  demonstration}},
  \href{https://doi.org/10.1103/PhysRevAccelBeams.23.072001}{\emph{Phys. Rev.
  Accel. Beams} {\bfseries 23} (2020) 072001}
  [\href{https://arxiv.org/abs/1807.03473}{{\ttfamily 1807.03473}}].

\bibitem{MICE:2019jkl}
{\scshape MICE} collaboration, \emph{{Demonstration of cooling by the Muon
  Ionization Cooling Experiment}},
  \href{https://doi.org/10.1038/s41586-020-1958-9}{\emph{Nature} {\bfseries
  578} (2020) 53} [\href{https://arxiv.org/abs/1907.08562}{{\ttfamily
  1907.08562}}].

\bibitem{mucollforum}
``Snowmass muon collider forum,.''
  \url{https://snowmass21.org/energy/muon_forum}.

\bibitem{nanni2021c}
M.~Bai, T.~Barklow, R.~Bartoldus, M.~Breidenbach, P.~Grenier, Z.~Huang et~al.,
  \emph{C~$^3$: {A "Cool" Route to the Higgs Boson and Beyond}},
  \href{https://arxiv.org/abs/2110.15800}{{\ttfamily 2110.15800}}.

\bibitem{C3demo}
R.~Patterson, M.~Liepe, J.~Maxson, S.~Belomestnykh, P.~Bhat, S.~Nagaitsev
  et~al., \emph{{C~$^3$ Demonstration Research and Development Plan}},  2022.

\bibitem{HELEN}
\emph{{H}iggs-{E}nergy {LE}pton ({HELEN}) {C}ollider based on advanced
  superconducting radio frequency technology: {R}eport to {S}nowmass 2021 (to
  be submitted)},  2022.

\bibitem{IOTA}
S.~Antipov, D.~Broemmelsiek, D.~Bruhwiler, D.~Edstrom, E.~Harms, V.~Lebedev
  et~al., \emph{{IOTA (Integrable Optics Test Accelerator): facility and
  experimental beam physics program}},
  \href{https://doi.org/10.1088/1748-0221/12/03/T03002}{\emph{JINST} {\bfseries
  12} (2017) T03002}.

\bibitem{Nagaitsev:2021xzy}
S.~Nagaitsev and V.~Lebedev, \emph{{A Cost-Effective Upgrade Path for the
  Fermilab Accelerator Complex}},
  \href{https://arxiv.org/abs/2111.06932}{{\ttfamily 2111.06932}}.

\bibitem{Belomestnykh:2021oyc}
S.A.~Belomestnykh, M.~Checchin, D.~Johnson, D.V.~Neuffer, S.E.~Posen,
  E.~Pozdeyev et~al., \emph{{An 8 GeV Linac as the Booster Replacement in the
  Fermilab Power Upgrade}},  \href{https://arxiv.org/abs/2203.05052}{{\ttfamily
  2203.05052}}.

\bibitem{Wang:2015xoa}
M.-H.~Wang, Y.~Nosochkov, Y.~Cai and M.~Palmer, \emph{{Design of a 6 TeV Muon
  Collider}},  in \emph{{6th International Particle Accelerator Conference}},
  p.~TUPTY081, 2015,
  \href{https://doi.org/10.18429/JACoW-IPAC2015-TUPTY081}{DOI}.

\bibitem{Neuffer:2013wrd}
D.~Neuffer, M.~Palmer, Y.~Alexahin, C.~Ankenbrandt and J.P.~Delahaye, \emph{{A
  muon collider as a Higgs factory}},  in \emph{{4th International Particle
  Accelerator Conference}}, 5, 2013
  [\href{https://arxiv.org/abs/1502.02042}{{\ttfamily 1502.02042}}].

\bibitem{IMCC}
``International muon collider collaboration,.''
  \url{https://muoncollider.web.cern.ch/}.

\bibitem{Kirk:2008zza}
H.G.~Kirk et~al., \emph{{The MERIT High-Power Target Experiment at the CERN
  PS}}, {\emph{Conf. Proc. C} {\bfseries 0806233} (2008) WEPP169}.

\bibitem{PhysRevLett.111.184802}
M.~Chung, M.G.~Collura, G.~Flanagan, B.~Freemire, P.M.~Hanlet, M.R.~Jana
  et~al., \emph{Pressurized ${\mathbf{h}}_{2}$ rf cavities in ionizing beams
  and magnetic fields},
  \href{https://doi.org/10.1103/PhysRevLett.111.184802}{\emph{Phys. Rev. Lett.}
  {\bfseries 111} (2013) 184802}.

\bibitem{PhysRevAccelBeams.20.032002}
K.~Yu, R.~Samulyak, K.~Yonehara and B.~Freemire, \emph{Simulation of
  beam-induced plasma in gas-filled rf cavities},
  \href{https://doi.org/10.1103/PhysRevAccelBeams.20.032002}{\emph{Phys. Rev.
  Accel. Beams} {\bfseries 20} (2017) 032002}.

\bibitem{Piekarz:2021mna}
H.~Piekarz, S.~Hays, B.~Claypool, M.~Kufer and V.~Shiltsev, \emph{{Record High
  Ramping Rates in HTS Based Super-conducting Accelerator Magnet}},  in
  \emph{{27th International Conference on Magnet Technology}}, 11, 2021
  [\href{https://arxiv.org/abs/2111.06459}{{\ttfamily 2111.06459}}].

\bibitem{Stratakis:2014nna}
D.~Stratakis and R.B.~Palmer, \emph{{Rectilinear six-dimensional ionization
  cooling channel for a muon collider: A theoretical and numerical study}},
  \href{https://doi.org/10.1103/PhysRevSTAB.18.031003}{\emph{Phys. Rev. ST
  Accel. Beams} {\bfseries 18} (2015) 031003}.

\bibitem{Bogacz:2017iia}
S.A.~Bogacz, \emph{{Muon Acceleration Concepts for NuMAX: ''Dual-use'' Linac
  and ''Dogbone'' RLA}},
  \href{https://doi.org/10.1088/1748-0221/13/02/P02002}{\emph{JINST} {\bfseries
  13} (2018) P02002} [\href{https://arxiv.org/abs/1708.01274}{{\ttfamily
  1708.01274}}].

\bibitem{PIC}
Y.~Derbenev, V.~Morozov, A.~Afanasev, K.~Beard, R.~Johnson, B.~Erdelyi et~al.,
  \emph{Parametric-resonance ionization cooling of muon beams}, {\emph{arXiv}
  }.

\bibitem{FOFOSnake}
Y.~Alexahin, \emph{Helical fofo snake for initial six-dimensional cooling of
  muons}, \href{https://doi.org/10.1088/1748-0221/13/08/P08013}{\emph{JINST}
  {\bfseries 13} (2018) P08013}.

\bibitem{Rogers:2021wyv}
{\scshape nuSTORM} collaboration, \emph{{nuSTORM Accelerator Challenges and
  Opportunities}},
  \href{https://doi.org/10.18429/JACoW-IPAC2021-THPAB175}{\emph{JACoW}
  {\bfseries IPAC2021} (2021) THPAB175}.

\bibitem{Brizzolari:2022seb}
C.~Brizzolari et~al., \emph{{The ENUBET monitored neutrino beam: a progress
  report}}, \href{https://doi.org/10.22323/1.398.0205}{\emph{PoS} {\bfseries
  EPS-HEP2021} (2022) 205}.

\bibitem{Adey:2017dvr}
D.~Adey, R.B.~Appleby, R.~Bayes, A.~Bogacz, A.D.~Bross, J.B.~Lagrange et~al.,
  \emph{{Overview of the Neutrinos from Stored Muons Facility - nuSTORM}},
  \href{https://doi.org/10.1088/1748-0221/12/07/P07020}{\emph{JINST} {\bfseries
  12} (2017) P07020}.

\bibitem{Lackowski:2013ria}
T.~Lackowski, S.~Dixon, R.~Jedziniak, M.~Blewitt and L.~Fink, \emph{{nuSTORM
  Project Definition Report}},
  \href{https://arxiv.org/abs/1309.1389}{{\ttfamily 1309.1389}}.

\bibitem{COMB}
H.~Kashikhin, V.~Lombardo and G.~Velev, \emph{{Magnet Design Optimization for
  Future Hadron Colliders}}, {\emph{Proc. of 10th International Particle
  Accelerator Conference (IPAC2019)} {\bfseries THPTS084} (2019)
  doi:10.18429/JACoW}.

\bibitem{CORC}
D.~van~der Laan, J.D.~Weiss and D.M.~McRaeand, \emph{{Status of CORC® cables
  and wires for use in high-field magnets and power systems a decade after
  their introduction}}, {\emph{2019 Supercond. Sci. Technol.} {\bfseries 32}
  (2019) 033001. doi:10.1088/1361}.

\bibitem{STAR}
S.~Kar, W.~Luo, A.B.~Yahia, X.~Li, G.~Majkic and V.~Selvamanickam,
  \emph{{Symmetric tape round REBCO wire with Je (4.2 K, 15 T) beyond 450 A
  mm−2 at 15 mm bend radius: a viable candidate for future compact
  accelerator magnet applications}}, {\emph{Supercond. Sci. Technol.}
  {\bfseries 31} (2018) 04LT01}.

\bibitem{Nb3SnFCC}
X.~Xu, X.~Peng, J.~Lee, J.~Rochester and M.~Sumption, \emph{{High Critical
  Current Density in Internally-oxidized Nb$_3$Sn Superconductors and its
  Origin}}, {\emph{Scr. Mater.} {\bfseries 186} (2020) 317}.

\bibitem{AI}
D.H.~et~al., \emph{{Intelliquench: An Adaptive Machine Learning System for
  Detection of Superconducting Magnet Quenches}}, {\emph{IEEE Trans. Applied
  Supercond.} {\bfseries 31} (2021) 1, doi: 10.1109/TASC.2021.3058229}.

\bibitem{MDP_WP}
\emph{A strategic approach to advanced magnet technology for the next
  generation colliders: {R}eport to {S}nowmass (to be submitted)},  2022.

\bibitem{FRMPiekarz}
H.~Piekarz, S.~Hays, J.~Blowers, B.~Claypool and V.~Shiltsev, \emph{{Record
  fast-cycling accelerator magnet based on HTS conductor}}, {\emph{Nucl. Instr.
  Methods} {\bfseries A943} (2019) 162490}.

\bibitem{FRMWhite}
\emph{Fast cycling {HTS}-based superconducting accelerator magnets: Feasibility
  study and readiness demonstration program driven by the neutrino physics and
  muon collider needs: {R}eport to {S}nowmass (to be submitted)},  2022.

\bibitem{LEAF}
\emph{White paper on {L}eading-{E}dge technology {A}nd {F}easibility-directed
  {P}rogram aimed at readiness demonstration for energy frontier circular
  colliders by the next decade: {R}eport to {S}nowmass (to be submitted)},
  2022.

\bibitem{2ndGEN}
\emph{Development and demonstration of next generation technology for
  {N}b$_3${S}n accelerator magnets with lower cost, improved performance
  uniformity, and higher operating point in the 12-14 t range: {R}eport to
  {S}nowmass (to be submitted)},  2022.

\bibitem{GARD-RF-Strategy}
\emph{{Radiofrequency Accelerator R\&D Strategy Report: DOE HEP general
  Accelerator R\&D RF Research Roadmap Workshop}},  2017.

\bibitem{SRFRDSnowmass2021_01}
S.~Posen et~al., \emph{{Key Directions for Research and Development of
  Superconducting Radiofrequency (SRF) Cavities}}, {\emph{Snowmass 2021 Letter
  of Intent, AF-206} (2021) }.

\bibitem{SRFRDSnowmass2021_02}
M.~Martinello et~al., \emph{{Plasma Processing for In-Situ Field Emission
  Mitigation of Superconducting Radiofrequency (SRF) Cryomodules}},
  {\emph{Snowmass 2021 Letter of Intent, AF-217} (2021) }.

\bibitem{SRFRDSnowmass2021_03}
R.~Geng et~al., \emph{{Field Emission Suppression in High-Gradient SRF Cavity
  Systems}}, {\emph{Snowmass 2021 Letter of Interest, AF-192} (2021) }.

\bibitem{SRFRDSnowmass2021_04}
A.~Gurevich et~al., \emph{{Challenges and opportunities of SRF theory for
  particle accelerators.}}, {\emph{Snowmass 2021 Letter of Interest, AF-122}
  (2021) }.

\bibitem{SRFRDSnowmass2021_05}
G.~Ciovati et~al., \emph{{SRF for future accelerators}}, {\emph{Snowmass 2021
  Letter of Intent, AF-171} (2021) }.

\bibitem{SRFRDSnowmass2021_06}
U.~Pudasaini et~al., \emph{{Next Generation SRF accelerators based on
  Nb$_3$Sn}}, {\emph{Snowmass 2021 Letter of Intent, AF-172} (2021) }.

\bibitem{SRFRDSnowmass2021_07}
A.-M.~Valente-Feliciano et~al., \emph{{Next-Generation Superconducting RF
  Technology based on Advanced Thin Film Technologies and Innovative Materials
  for Accelerator Enhanced Performance \& Energy Reach}}, {\emph{Snowmass 2021
  Letter of Intent, AF-205} (2021) }.

\bibitem{SRFRDSnowmass2021_08}
S.~Posen et~al., \emph{{Key Directions for Research and Development of
  Superconducting Radiofrequency (SRF) Cavities}}, {\emph{Snowmass 2021 Letter
  of Intent, AF-206} (2021) }.

\bibitem{SRFRDSnowmass2021_09}
M.~Checchin et~al., \emph{{Innovative Materials and Surface Treatments for SRF
  applications}}, {\emph{Snowmass 2021 Letter of Intent, AF-173} (2021) }.

\bibitem{SRFRDSnowmass2021_10}
S.~Balachandran et~al., \emph{{The necessity of a basic materials research
  community for the accelerated development of SRF materials}}, {\emph{Snowmass
  2021 Letter of Intent, AF-229} (2021) }.

\bibitem{SRFRDSnowmass2021_11}
E.~Barzi et~al., \emph{{SRF - An Impartial Perspective for Superconducting
  Nb$_3$Sn coated Copper RF Cavities for Future Linear Accelerators}},
  {\emph{Snowmass 2021 Letter of Intent, AF-248} (2021) }.

\bibitem{SRFRDSnowmass2021_12}
G.~Myneni et~al., \emph{{SRF - Development of High-efficiency and
  Cost-effective Forged Ingot Niobium Technology for Science Frontiers and
  Accelerator Applications}}, {\emph{Snowmass 2021 Letter of Intent, AF-087}
  (2021) }.

\bibitem{SRFRDSnowmass2021_13}
T.~Tajima et~al., \emph{{Development of MgB$_2$ Coated Superconducting
  Cavities}}, {\emph{Snowmass 2021 Letter of Intent, AF-057} (2021) }.

\bibitem{SRFRDSnowmass2021_14}
S.~Posen et~al., \emph{{Nb$_3$Sn Superconducting Radiofrequency Cavities}},
  {\emph{Snowmass 2021 Letter of Intent, AF-086} (2021) }.

\bibitem{Belomestnykh2012}
S.~Belomestnykh, \emph{{Superconducting radio-frequency systems for
  high-$\beta$ accelerators}},
  \href{https://doi.org/10.1142/S179362681230006X}{\emph{Rev. Accel. Sci.
  Technol.} {\bfseries 5} (2012) 147}.

\bibitem{CCCdemonstrator}
\emph{{$C^3$} {D}emonstration {R}esearch and {D}evelopment {P}lan: {R}eport to
  {S}nowmass 2021 (to be submitted)},  2022.

\bibitem{lbnf2016}
V.~Papadimitrious et~al., \emph{{Design of the LBNF Beamline}},
  {\emph{Fermilab-Conf-17-022-AD} (2016) }.

\bibitem{jparc2017}
M.~Friends et~al., \emph{J-parc accelerator and neutrino beamline update
  programme}, {\emph{IOP Conf.} {\bfseries 888} (2017) 012042}.

\bibitem{mcJP}
J.~Delahaye et~al., \emph{A staged muon accelerator facility for neutrino and
  colllider physics}, {\emph{arXiv} 1502.01647}.

\bibitem{fcccdr}
``{FCC CDR}.'' \url{fcc-cdr.web.cern.ch}.

\bibitem{Frederique2021}
F.~Pellemoine et~al., \emph{Novel materials to improve high power target
  reliability}, {\emph{Snowmass LOI} (2022) }.

\bibitem{Kavin2021}
K.~Ammigan et~al., \emph{Status and update of the radiate collaboration r\&d},
  {\emph{Proc. Acc. App.} {\bfseries 17} }.

\bibitem{Ishida2021}
T.~Ishida et~al., \emph{Tensile behavior of dual-phase titanium alloys under
  high-intensity proton beam exposure: radiation-induced omega phase
  transformation in ti-6al-4v}, {\emph{arXiv} 2004.11562}.

\bibitem{Atwani2021}
O.~El-Atwani et~al., \emph{Outstanding radiation resistance of tungsten-based
  high entropy alloys},
  \href{https://doi.org/10.1126/sciadv.aav2002}{\emph{Sci. Adv.} {\bfseries 5}
  (2019) eaav200}.

\bibitem{Sujit2021}
S.~Bidhar, \emph{{Electrospun nanofiber materials for high power target
  applications}}, {\emph{Fermilab-slides-17-017-AD} (2017) }.

\bibitem{BRN_Detectors}
B.~Fleming et~al., \emph{{Basic Research Needs for High Energy Physics Detector
  Research \& Development: Report of the Office of Science Workshop on Basic
  Research Needs for HEP Detector Research and Development: December 11-14,
  2019}},  12.
\newblock https://doi.org/10.2172/1659761.

\bibitem{ECFA_Detectors}
{ECFA Detector R\&D Roadmap Process Group}, \emph{{The 2021 ECFA detector
  research and development roadmap}},  Tech. Rep.
  \href{https://cds.cern.ch/record/2784893}{CERN-ESU-017}, Geneva (2020),
  \href{https://doi.org/10.17181/CERN.XDPL.W2EX}{DOI}.

\bibitem{IFFacilities}
\emph{White paper on test beam and irradiation facilities for detector research
  and development: {R}eport to {S}nowmass (to be submitted)},  2022.

\end{thebibliography}\endgroup
\end{document}